\documentclass[twocolumn,showpacs,preprintnumbers,amsmath,amssymb]{revtex4}

\usepackage{graphicx}
\usepackage{dcolumn}
\usepackage{bm}

\begin{document}

\preprint{APS/123-QED}

\title{Stellar (n,$\gamma$) cross sections of $p$-process isotopes \\
Part 1: $^{102}$Pd, $^{120}$Te, $^{130,132}$Ba, and $^{156}$Dy}

\author{I. Dillmann}\email{iris.dillmann@ph.tum.de}
 \altaffiliation[Previous addresses: ]{Departement f\"ur Physik, Universit\"at Basel and Institut f\"ur Kernphysik, Forschungszentrum Karlsruhe, D-76021 Karlsruhe.}
 \affiliation{Physik Department E12 and Excellence Cluster Universe, Technische Universit\"at M\"unchen, Germany}

\author{C. Domingo-Pardo}
	\altaffiliation[Previous address: ]{Institut f\"ur Kernphysik, Forschungszentrum Karlsruhe, D-76021 Karlsruhe.}
\author{M. Heil}
  \altaffiliation[Previous address: ]{Institut f\"ur Kernphysik, Forschungszentrum Karlsruhe, D-76021 Karlsruhe.}
 \affiliation{Gesellschaft f\"ur Schwerionenforschung Darmstadt, Germany}

\author{F. K\"appeler and S. Walter}
 \affiliation{Institut f\"ur Kernphysik, Forschungszentrum Karlsruhe, Postfach 3640, D-76021 Karlsruhe, Germany}

\author{S. Dababneh}
 \affiliation{Faculty of Applied Sciences, Al-Balqa Applied University, Salt 19117, Jordan}

\author{T. Rauscher}
\author{F.-K. Thielemann}
 \affiliation{Departement Physik, Universit\"at Basel, Klingelbergstrasse 82, CH-4056 Basel, Switzerland}

\date{\today}

\begin{abstract}
We have investigated the $(n,\gamma)$ 
cross sections of $p$-process isotopes with the activation technique. 
The measurements were carried out at the Karlsruhe Van de Graaff 
accelerator using the $^7$Li$(p,n)$$^7$Be source for simulating a 
Maxwellian neutron distribution of $kT$= 25 keV. Stellar cross section measurements are reported for the light 
$p$-process isotopes $^{102}$Pd, $^{120}$Te, $^{130,132}$Ba, and 
$^{156}$Dy. In a following paper the cross sections of $^{168}$Yb, 
$^{180}$W, $^{184}$Os, $^{190}$Pt, and $^{196}$Hg will be discussed. 
The data are extrapolated to $p$-process energies by including
information from evaluated nuclear data libraries. The results are compared to standard
Hauser-Feshbach models frequently used in astrophysics.
\end{abstract}

\maketitle

\section{\label{Intro}Introduction}
Astrophysical models can explain the origin of most nuclei beyond
the iron group by a combination of processes involving neutron
captures on long ($s$-process) or short ($r$-process)
time scales \cite{bbfh57, lawi01}. 

\begin{figure}[!htb]
\includegraphics[scale=0.60]{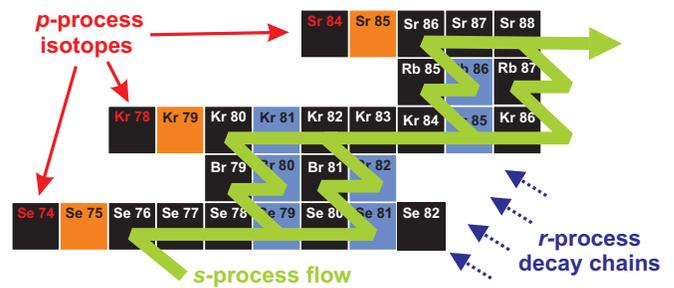}
\caption{\label{p-iso} (Color online) Example from the chart of 
nuclides illustrating the position of the lightest $p$-nuclei, 
which are shielded from the $s$-process flow and the $r$-process 
decay chains.}
\end{figure}

However, 32 proton-rich stable isotopes between $^{74}$Se and
$^{196}$Hg cannot be formed in these neutron capture processes, 
because they are either shielded by stable isotopes from the 
$r$-process decay chains or lie outside the $s$-process flow 
(Fig.~\ref{p-iso}). These isotopes, which are ascribed to the 
so-called "$p$-process", are 10 to 100 times less abundant 
than their $s$- and $r$-process neighbors. So far, the 
astrophysical site of the $p$-process is still under discussion, 
since the solar $p$-abundances can not be completely described 
by current models. 

Historically, the $p$-process was thought to proceed via proton 
captures, but a plausible site with the required amount of 
free protons could not be identified. Moreover, elements with large $Z$
cannot be produced by proton captures because the temperatures necessary
to overcome the Coulomb repulsion favor photodisintegration rather than
charged-particle capture.

The most plausible astrophysical site is the explosively burning 
Ne/O layer in core collapse supernovae, which is heated to
ignition temperatures by the outgoing shock front 
\cite{woho78,woho90,raar95}. In this high-temperature environment
proton-rich nuclei are produced by sequences of photo-dissociations 
and $\beta^+$ decays. In stars 20 times more massive than the sun
the $p$-process temperatures for efficient photo-disintegration 
are already reached at the end of hydrostatic Ne/O burning 
\cite{rahe02}. This mechanism is also called "$\gamma$ process" because proton-rich 
isotopes are produced by ($\gamma, n$) reactions on pre-existing 
seed nuclei from the $s$- and $r$- processes. When ($\gamma, p$) and 
($\gamma, \alpha$) reactions become comparable or faster than 
($\gamma, n$), the reaction path branches out from the initial 
isotopic chain and feeds nuclei with lower atomic number $Z$. 
While photodisintegration dominates in the early, hot phase, the
initially released neutrons can be recaptured at a later time, when
the material cools down after the passage of the shockwave.
The typical $p$-process abundance pattern exhibits
maxima at $^{92}$Mo ($N$=50) and $^{144}$Sm ($N$=82).

The solar abundances of the $p$-nuclei are reproduced
by current models of the $\gamma$ process within factors of two 
to three on average \cite{raar95,ray90,rahe02}, except for two
regions with nuclei of $A<100$ and $150\leq A\leq 165$ \cite{rahe02}.
The most abundant $p$-isotopes, $^{92,94}$Mo and $^{96,98}$Ru, are 
significantly underproduced because appropriately abundant seed nuclei are
missing. Alternative processes and sites have 
been proposed in order to explain this deficiency, i.e.\ reactions 
induced by the strong neutrino fluxes in the deepest ejected 
layers of core-collapse supernovae (the $\nu p$ process \cite{FML06}),
or explosive hydrogen burning in proton-rich, hot matter accreted onto the 
surface of neutron stars (the $rp$ process \cite{scha98,scha01}).
An alternative site for additional production of the $150\leq A\leq 165$
has not been suggested so far.

A few $p$-nuclides may also be produced by neutrino reactions 
during the $\gamma$ process. This "$\nu$ process" \cite{WHH90} 
could be the origin of the odd-odd isotopes $^{138}$La and 
$^{180}$Ta$^{\rm m}$, which are strongly underproduced in the 
$\gamma$ process. The abundances of both $p$-nuclei may be 
explained by neutrino scattering on their abundant neighbor
isotopes to states above the neutron emission threshold. 

The isotopes $^{152}$Gd, $^{164}$Er, and $^{180}$Ta$^{\rm m}$
were sometimes also considered as $p$-nuclei but it was 
found that significant fractions are produced indeed by
the $s$-process \cite{argo03}.

The fact that self-consistent studies of the $\gamma$ process 
have problems to synthesize the $p$-nuclei in the mass regions 
$A<124$ and $150\leq A\leq 165$ \cite{rahe02} may result from 
difficulties related to the astrophysical models as well as from systematic
uncertainties of the nuclear physics input. Therefore, the 
improvement of nuclear reaction cross sections is crucial for 
further progress in $p$-process models, either by directly 
replacing theoretical predictions by experimental data or by 
testing the reliability of predictions if the relevant energy 
range is not accessible by experiments. 

In this context we have carried out an extensive experimental 
program to measure the ($n, \gamma$) cross sections of 13 $p$-only 
isotopes by means of the activation technique. Two publications 
are already available concerning $^{74}$Se, $^{84}$Sr \cite{DHK06},
and $^{174}$Hf \cite{VDH07}. The present paper continues this 
series of measurements with the isotopes $^{102}$Pd, $^{120}$Te, 
$^{130,132}$Ba, and $^{156}$Dy, and a follow-up paper will cover 
the remaining heavy $p$-isotopes $^{168}$Yb, $^{180}$W, $^{184}$Os, 
$^{190}$Pt, and $^{196}$Hg. A concluding paper will present $p$-process network 
calculations based on a new version of the "Karlsruhe Astrophysical 
Database of Nucleosynthesis in Stars" ({\sc KADoNiS}) \cite{kado06},
where the available experimental and semi-empirical ($n, \gamma$) 
cross sections for the $p$-process will be added to the already 
existing data library for the $s$-process. Thereby, the 
{\sc KADoNiS} project will be extended to provide the $p$-process 
community with updated experimental information. These data 
will necessarily remain a complement to the indispensable 
theoretical predictions for the vast majority of the mostly 
unstable isotopes in the $p$-process network, which are not
accessible to cross section measurements with present experimental 
techniques. Nevertheless, these data provide important tests of 
existing calculations in the Hauser-Feshbach statistical model 
\cite{hafe52}, i.e. with the codes {\sc NON-SMOKER} 
\cite{rath00,rath01} or {\sc MOST} \cite{most05}.

The experimental aspects of this work are presented in Sec.~\ref{Exp}.
The analysis of the data and a discussion of the related uncertainties 
follows in Secs.~\ref{Data} and \ref{Unc}. The results are discussed in
Sec.~\ref{Res} and the calculated Maxwellian averaged cross sections 
in Sec.~\ref{macs}. 

\section{\label{Exp}Experimental technique}
This section contains a concise discussion of the experimental 
technique. More detailed information can be found in Refs. 
\cite{beer80,raty88}.

\subsection{Neutron activation}
\begin{figure*}[!htb]
\includegraphics[scale=1]{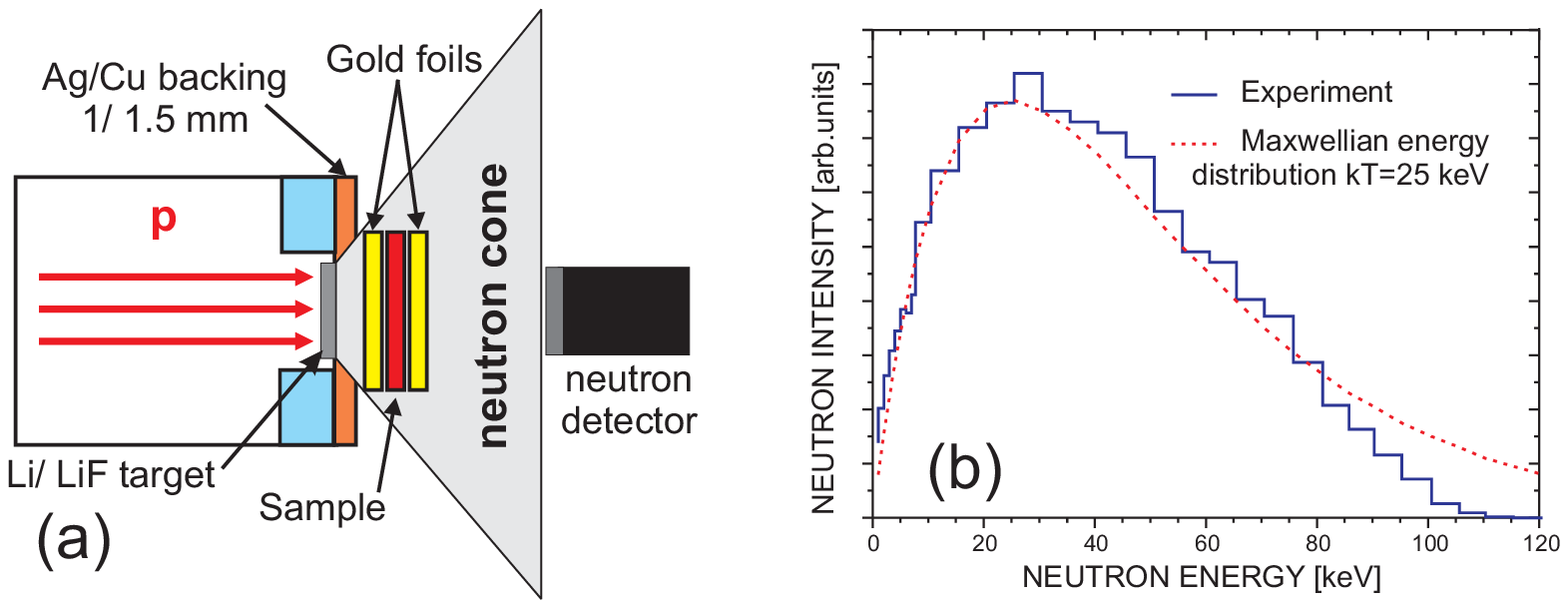}
\caption{\label{target}(Color Online) (a) Schematic sketch 
of the experimental setup. (b) Comparison of the experimental 
neutron distribution and a Maxwell distribution of $kT$=25~keV.}
\end{figure*}

The present capture measurements were carried out at the (now closed) Karlsruhe
3.7 MV Van de Graaff accelerator using the activation technique.
Neutrons were produced with the $^7$Li($p, n$)$^7$Be source by
bombarding 30 $\mu$m thick layers of metallic Li or crystalline
LiF on water-cooled Cu backings with protons of 1912 keV, 31 keV
above the threshold of the $^7$Li($p, n$)$^7$Be reaction at 1881~keV. 
Under these conditions, all neutrons are emitted into a forward 
cone of 120$^\circ$ opening angle. The resulting neutron field 
represents a quasi-stellar spectrum, which approximates a 
Maxwell-Boltzmann distribution for $kT$= \mbox{25.0 $\pm$ 0.5 keV} 
\cite{raty88} (see Fig.~\ref{target}). Activation in this spectrum yields, therefore, 
directly the Maxwellian averaged cross section (MACS), with only 
a small correction for the fact that the quasi-stellar spectrum 
is truncated at $E_n$= 106~keV. Neutron scattering in the Cu 
backing is negligible, since the transmission is $\approx$98\% 
in the energy range of interest. 

In all cases, the sample material was of natural composition and high elemental purity ($\geq 99.5\%$), either in metallic form or as a compound (Table~\ref{iso}). However, it has to be emphasized that the large uncertainty of the $^{120}$Te abundance given in \cite{iupac03} originates from the fractionation of tellurium in various materials and the fact that up to now no absolute isotopic abundance measurement has been carried out. 

The recommended value of 0.09 (1)\% in \cite{iupac03} is commented with "An electron multiplier was used for these measurements and the measured abundances were adjusted using a 'square root of the masses' correction factor" \cite{SRL78}. Independent relative measurements of the isotope ratios have been carried out later by De Laeter et al. \cite{deLae94} with a Faraday Cup collector and Lee et al. \cite{Lee95} using ICP-MS (inductively coupled plasma mass spectrometry). Both methods revealed $^{120}$Te abundances of 0.0918 (7)\% and 0.0927 (4)\%, respectively, which are in clear disagreement with the "best value" assignment of 0.0960 (7)\% by Smith et al. \cite{SRL78} given in the latest IUPAC Technical Report 2003 \cite{iupac03}. In \cite{deLae94} the value from Smith et al. \cite{SRL78} was corrected with the above mentioned factor $\sqrt{m_1/m_2}$ and one can calculate a value of 0.0935 (7)\% from the given isotope ratios. The correction is due to a mass discrimination which occurs when using electron multipliers. The signal from light isotopes is enhanced compared to that from heavier isotopes because the secondary electron yield at the first dynode is velocity dependent \cite{deLae94}. The methods using Faraday cups \cite{deLae94} or ICP-MS \cite{Lee95} do already account for this. From these three values we can derive a weighted average of 0.0927 (9)\% where the uncertainty is derived from the standard deviation of the measurements. We decided to use this value for our measurement instead of the "representative isotopic composition" given in \cite{iupac03}.

\begin{table}[!htb]
\caption{Sample materials and isotopic abundances \cite{iupac03}
\label{iso}}.  
\begin{ruledtabular}
\begin{tabular}{cccc}
Element & Isotope       & Sample material & Rel. abundance~[\%] \\
\hline
Pd      & $^{102}$Pd    & Pd (metal)      & 1.02 (1)            \\
Te      & $^{120}$Te    & Te (metal)      & 0.09 (1)$^a$		\\
Ba      & $^{130}$Ba    & BaCO$_3$        & 0.106 (1)           \\
        & $^{132}$Ba    & BaCO$_3$        & 0.101 (1)           \\
Dy		  & $^{156}$Dy	  & Dy (metal)      &	0.056 (3)	\\       
\end{tabular}
\end{ruledtabular}
$^a$ See text. Used value is 0.0927 (9)\%.
\end{table}

Apart from the Pd samples, which were cut from 25 $\mu$m thick foils, thin 
pellets 6 to 10 mm in diameter were pressed from the respective 
powder or granules and enclosed in thin cannings made from 
15 $\mu$m thick aluminum foil. During the irradiations the 
samples were sandwiched between 10-30 $\mu$m thick gold foils 
of the same diameter. In this way the neutron flux can be 
determined relative to the well-known capture cross section 
of $^{197}$Au \cite{raty88}. 

The activation measurements were carried out with the Van de Graaff 
accelerator operated in DC mode with a current of $\approx$100~$\mu$A 
(for the Li targets) or even higher currents (up to 150~$\mu$A) for 
the LiF targets. To ensure homogeneous illumination of the entire 
surface, the proton beam was continuously wobbled across the Li target. 
The samples were irradiated in close contact with the Li target 
with average neutron intensities of (1.5--3)$\times$10$^9$ s$^{-1}$ at 
the position of the samples. The neutron intensity was recorded in 
intervals of 60 or 90~s using a $^6$Li-glass detector 91 cm downstream 
of the lithium target. With this information, fluctuations in the 
neutron yield could be properly considered in the later correction 
of the number of nuclei, which decayed during the activation.

Over the course of the present measurements, several independent 
activations have been carried out for each isotope with modified 
experimental parameters (see Table~\ref{tab:act}).

\begin{table}[!htb]
\caption{\label{tab:act}Sample characteristics and activation parameters.}
\renewcommand{\arraystretch}{1.1} 
\begin{ruledtabular}
\begin{tabular}{ccccccc}
Target  & Sample  & Diameter & \multicolumn{2}{c}{Mass}  & $t_{a}$ & $\Phi_{tot}$$^a$ \\
\cline{4-5}
isotope &         & [mm]     & [mg] & [10$^{18}$ atoms] & [min] 	& [10$^{14}$ n] \\
\hline
$^{102}$Pd & pd-1 & 10       & 452.5 & 26.1       & 9770    & 8.18		  \\
           & pd-2 & 8        & 301.5 & 17.4 	  & 5751    & 4.83		  \\
           & pd-3 & 12       & 339.5 & 19.6 	  & 7585    & 3.48		  \\
		   & 	  &			 & 		 & 			  &			& 			  \\
$^{120}$Te & te-1 & 10    	 & 352.9 & 1.54 	  & 2617   & 1.96		  \\
    	   & te-2 & 10    	 & 441.2 & 1.93 	  & 1600   & 1.52		  \\
    	   & te-3 & 8     	 & 349.3 & 1.53 	  & 1406   & 1.56		  \\
    	   & te-4 & 8     	 & 417.2 & 1.83 	  & 4142   & 3.03		  \\
    	   & te-5 & 8     	 & 409.6 & 1.79 	  & 2593   & 3.09		  \\
		   & 	  &			 & 		 & 			  &		   & 			  \\
$^{130}$Ba & ba-1 & 8 		 & 106.9 & 0.346 	  &  7721  & 6.93 		  \\
$^{132}$Ba & 	  & 	     &		 & 0.330 	  &		   &			  \\				
$^{130}$Ba & ba-2 & 8 		 & 145.5 & 0.471 	  &  4014  & 2.70		  \\
$^{132}$Ba & 				 &		 & & 0.448 	  &		   &			  \\
$^{130}$Ba & ba-3 & 10		 & 149.7 & 0.484 	  &  4280  & 4.48		  \\
$^{132}$Ba & 				 &		 & & 0.461 	  &		   &			  \\	
		   & 	  &			 &		 & 		  	  & 	   & 			  \\
$^{156}$Dy & dy-1 & 6  		 & 28.3  & 0.0588 	  & 964   & 0.995 		  \\
	 	   & dy-2 & 10 		 & 80.1  & 0.166 	  & 362   & 0.416 		  \\
	 	   & dy-3 & 6  		 & 61.6  & 0.128 	  & 902   & 1.73 		  \\
\end{tabular}
\end{ruledtabular}
$^a$ Total neutron exposure during activation.
\end{table}

\subsection{Activity measurements}

\begin{figure}[!htb]
\includegraphics[scale=1]{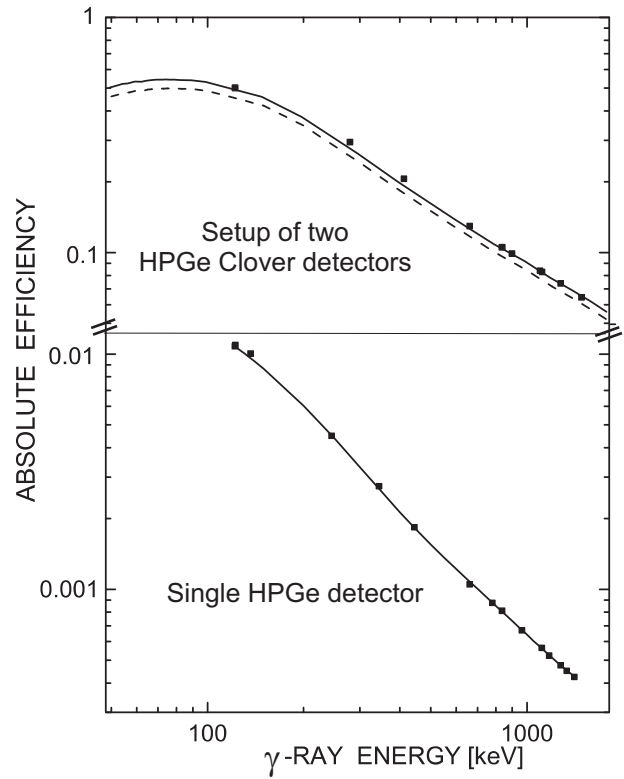}
\caption{\label{effi} Efficiency curves of the Clover detector 
system and the single HPGe detector. The statistical uncertainties
of the calibration measurements correspond to the size of the 
symbols. The simulated efficiency of the Clover system (dashed 
line) was normalized to fit the measured data points.}
\end{figure}

For the measurement of the induced activities two detector setups 
were available. A single high purity Germanium (HPGe) detector 
with a well defined measuring position at 76.0$\pm$0.5~mm from 
the detector surface was used for counting the activities of the gold
foils and of the $^{121}$Te, $^{131}$Ba, and $^{133}$Ba$^{\rm m}$ decays. 
The detector was shielded by 10~cm of lead and 5 mm of copper. Energy 
and efficiency calibrations have been carried out with a set of
reference $\gamma$-sources in the energy range between 60 keV and
2000 keV (see Fig.~\ref{effi}).

The small $\gamma$ activities of $^{103}$Pd and $^{133}$Ba$^{\rm g}$ 
were measured with a system of two HPGe Clover detectors 
(see Ref.~\cite{clover} for more details). Each Clover detector
consists of four independent HPGe n-type crystals in a common 
cryostat. The two Clovers were placed face to face, in contact 
with a 5.2~mm thick sample holder, corresponding nearly to a 
4$\pi$ geometry. The sample position in the very 
center of the system could be reproduced within $\pm$0.1 mm. The 
whole assembly was shielded against room background with 10~cm of 
lead and a 5 mm thick layer of copper. The data from the eight Ge 
crystals of the two Clover detectors were processed by separate 
analog to digital converters (ADCs) and could, therefore be
analyzed independently. In Clover measurements, the contributions 
of the eight crystals were added to represent the total number of 
events per $\gamma$-ray line. The efficiency calibration of the 
Clover system was carried out with a set of weak reference sources.

\section{\label{Data}Data analysis}

\subsection{General procedure}

The total amount of activated nuclei $N_{act}$ at the end of each
irradiation can be deduced from the number of events $C$ in a
particular $\gamma$-ray line registered in the HPGe detector
during the measuring time $t_m$ \cite{beer80},
\begin{eqnarray}
N_{act} = \frac{C(t_m)} {\varepsilon_\gamma~ I_\gamma~ K~
(1-{\rm e}^{-\lambda~t_m})~{\rm e}^{-\lambda~t_w}} \label{eq:Z}
\end{eqnarray}
where $t_w$ denotes the waiting time between irradiation and
activity measurement, $\varepsilon_\gamma$ the efficiency of
the HPGe detector, and $I_\gamma$ the relative $\gamma$ intensity 
per decay of the respective transition. $K$ is a correction factor which is either the 
self-absorption correction factor $K_\gamma$ (Eq~\ref{eq:K}) or the total correction factor $K_{tot}$ (Table~\ref{geant}). The decay properties of the investigated product nuclei are summarized in 
Table~\ref{tab:decay}.

\begin{table*}[!htb]
\caption{\label{tab:decay}Decay properties of the product nuclei used in the analysis. EC stands for electron capture decay, IT for isomeric transition.}
\renewcommand{\arraystretch}{1.2} 
\begin{ruledtabular}
\begin{tabular}{ccccccc}
Product             & t$_{1/2}$     & Decay mode        & E$_\gamma$ & I$_\gamma$ & Reference     \\
 nucleus            &               &                   & [keV]      & [\%]       & 		      \\
\hline
$^{103}$Pd          & 16.991 (19)~d & EC 	  			& 357.5      & 0.0221 (7) & \cite{nds103} \\
			        &				   & 	            & 			 & 		   &   				  \\
$^{121}$Te$^{\rm g}$& 19.16 (5)~d   & EC 				& 573.1      & 80.3 (25)  & \cite{nds121} \\
$^{121}$Te$^{\rm m}$& 154 (7)~d     & IT (88.6 (11) \%) & 212.2      & 81.4 (1)   & 		  	  \\
$^{121}$Te$^{\rm m}$&               & EC (11.4 (11) \%) & 1102.1     & 2.54 (6)   & 			  \\
					&			   	&					& 	 		 & 			  & 		      \\
$^{131}$Ba          & 11.50 (6)~d   & EC 				& 216.1 	 & 20.4 (4) & \cite{nds131new}\\
                 		&           		& EC 				& 496.3 	 & 48.0 (4)   & 			  \\
			        &				& 					& 			 & 		   	  &				  \\
$^{133}$Ba$^{\rm g}$& 10.52 (13) yr & EC 				& 356.0 	 & 62.05 (19) & \cite{nds133a}\\
$^{133}$Ba$^{\rm m}$& 38.9 (1)~h  	& IT (99.99 \%) 	& 275.9 	 & 17.8 (6)   &               \\
			        &				& 	            	& 			 & 		   	  &   			  \\
$^{157}$Dy  		&  8.14 (4)~h 	& EC 				& 326.3  	 & 93 (3) 	  & \cite{nds157} \\
			        &				& 	            	& 			 & 		   	  &   			  \\
$^{198}$Au  		& 2.69517 (21)~d & $\beta$$^-$ 		& 411.8 	 & 95.58 (12) & \cite{nds198} \\
\end{tabular}
\end{ruledtabular}
\end{table*}

The large distance between sample and detector in the measurements 
with the single HPGe detector allowed us to calculate the correction 
for $\gamma$-ray self-absorption from the expression for disk samples 
of thickness $d$ \cite{beer80}, 
\begin{eqnarray}
K_\gamma = \frac{1-{\rm e}^{-\mu d}}{\mu d}, \label{eq:K}
\end{eqnarray}
using the $\gamma$-ray absorption coefficients $\mu$ from 
Ref.~\cite{nist}. This correction factor was negligible
for the thin (10-30~$\mu$m) gold foils. 

The close geometry of the Clover detectors required a 
more elaborate treatment of the sample-related
corrections in the measurement of the small activities 
of $^{103}$Pd and $^{133}$Ba$^{\rm g}$. The correction 
factors $K'_\gamma$ for $\gamma$-ray self absorption, 
$K_E$ for the extended geometry of the sample, 
and $K_S$ for the summing effect of cascade transitions
have the been calculated by means of Monte Carlo 
simulations with the GEANT4 toolkit \cite{Ago03} and a 
detailed computer model of the setup \cite{clover}
(Table~\ref{geant}).
The summing correction factor $K_S$ of sample pd-3 is slightly higher than for samples pd-1 and pd-2 due to the varying thickness. The thinner sample pd-3 absorbs less X-rays and thus more coincident summing with decay $\gamma$-rays can occur.
 
\begin{table}[!htb]
\caption{GEANT4 simulations of the correction factors for the 
$^{103}$Pd and $^{133}$Ba$^{\rm g}$ measurements with the Clover 
detector system.}\label{geant}
\begin{ruledtabular}
\begin{tabular}{cccccc}
Sample & Thickness & $K_E$ 	& $K_\gamma'$ & $K_S$  & $K_{tot}$ \\
	   & [mm] 	   & 	   	& 			 & 		  & 			\\
\hline
pd-1 & 0.5 		   & 0.9976 & 0.9563 	 & 0.9991 & 0.9531 		\\
pd-2 & 0.5 		   & 0.9986 & 0.9569 	 & 0.9951 & 0.9509 		\\
pd-3 & 0.25 	   & 0.9972 & 0.9762 	 & 0.9663 & 0.9407 		\\
	 & 			   & 		& 		  	 &		  & 			\\
ba-1 & 0.48 	   & 1.0002 & 0.9820 	 & 0.9575 & 0.9405 		\\
ba-2 & 0.65 	   & 1.0022 & 0.9752 	 & 0.9608 & 0.9390 		\\
ba-3 & 0.43 	   & 0.9994 & 0.9841 	 & 0.9570 & 0.9412 		\\
\end{tabular}
\end{ruledtabular}
\end{table} 

The number of activated nuclei $N_{act}$ can be written as
\begin{eqnarray}
N_{act}(i) = \langle \sigma_i \rangle~N_i~\Phi_{tot}~f(i) \quad, \label{eq:5}
\end{eqnarray}
where $\Phi_\mathrm{tot} = \int \phi(t)dt$ is the time-integrated neutron
flux and $N_i$ the number of atoms of species $i$ in the sample. As our
measurements are carried out relative to $^{197}$Au as a standard,
the neutron flux $\Phi_{tot}$ cancels out in the ratio,

\begin{eqnarray}
\frac{N_\mathrm{act}(i)}{N_\mathrm{act}(\mathrm{Au})}=
\frac{\langle \sigma_i \rangle~N_i~f(i)}{\langle \sigma_\mathrm{Au} \rangle~N_\mathrm{Au}~f(\mathrm{Au})} \nonumber\\
\Longleftrightarrow  \langle \sigma_i \rangle = \langle \sigma_\mathrm{Au} \rangle
\frac{N_\mathrm{act}(i)~N_\mathrm{Au}~f(\mathrm{Au})}{N_\mathrm{act}(\mathrm{Au})~N_i~f(i)} \quad.
\label{eq:6}
\end{eqnarray}

The correction factor
\begin{eqnarray}
f =
\frac{\int_{0}^{t_a}\phi(t)~{\rm e}^{-\lambda(t_a-t)}~dt}{\int_{0}^{t_a}\phi(t)~dt}
\label{eq:fb}
\end{eqnarray}
for the decay of activated nuclei during the irradiation time 
$t_a$ is calculated from the neutron flux history recorded 
with the $^6$Li glass detector downstream of the neutron 
target. This correction includes also the effect of variations 
in the neutron flux. 

The cross section $\langle \sigma_{exp} \rangle$ is given in brackets to indicate that it represents an average over the 
quasi-stellar spectrum of the $^{7}$Li(p,n)$^{7}$Be source.
The reference value for the experimental averaged $^{197}$Au cross 
section was adopted as $\langle \sigma_\mathrm{exp} \rangle$=$586 \pm 8$ mbarn \cite{raty88}. 

\subsection{Partial cross sections}

In the activations of $^{120}$Te, $^{130}$Ba, and $^{132}$Ba, 
neutron capture populates ground and isomeric states in the 
product nucleus. The partial cross section to $^{131}$Ba$^{\rm m}$
could not be measured in this work because the isomer is too
short short-lived ($t_{1/2}=14.6$ min). Therefore, only the 
total capture cross section of $^{130}$Ba was derived from 
the ground state activity after an appropriate waiting time. 
Similarly, the isomeric state in $^{133}$Ba decays with 99.99\% 
probability by internal transitions with a half-life of 38.9~h 
so that the total cross section can later be derived from the 
ground-state activity ($t_{1/2}=10.52$~yr). In this case, the
isomer lived long enough that the partial cross section could
be determined as well. 

In cases where the half-lives of ground state and 
isomer are of the same order of magnitude, as for $^{121}$Te,
where the isomer (88.6\% IT, 11.4\% EC) is even longer-lived
than the ground-state, the contributions of both states to 
the total cross section have to be properly disentangled.
The partial cross section to the ground state can
be deduced from the $\gamma$ spectra of the first few days, 
where the contribution from the isomer decay is still small. 
The exact correction for the ground-state was 
discussed in detail in Ref.~\cite{DHK06} and has been used here.

\section{Uncertainties}\label{Unc}

The experimental uncertainties are summarized in Table~\ref{tab:err}. 
Since nearly every stellar neutron cross section measurement 
was carried out relative to gold, the  1.4\% uncertainty of the 
gold cross section \cite{raty88} cancels out in most astrophysical
applications. 

\begin{table*}[!htb]
\caption{Compilation of uncertainties.}\label{tab:err}
\renewcommand{\arraystretch}{1.2} 
\begin{ruledtabular}
\begin{tabular}{lcccccccc}
Source of uncertainty & \multicolumn{8}{c}{Uncertainty (\%)}\\
  & $^{197}$Au &$^{102}$Pd & $^{120}$Te$\rightarrow$g & $^{120}$Te$\rightarrow$m 
	& $^{130}$Ba $^a$ & $^{132}$Ba$\rightarrow$m & $^{132}$Ba$\rightarrow$g+m & $^{156}$Dy \\
\hline
Gold cross section           & 1.4$^b$	 &	--       & --        & --        & --        & --        & --        & --       \\		
Isotopic abundance    		 & -- 		 & 1.0       & 0.9       & 0.9       	& 0.9       & 1.0       & 1.0       & 5.4 		\\
Detector efficiency   		 & 2.0 		 & \multicolumn{7}{c}{2.0}   	 		   		   			   			   			\\
Divergence of flux    		 & --  		 & \multicolumn{7}{c}{2.0} 				   	 										\\
Sample mass                  & 0.2 		 & --        & --        & --        & --        & --        & --        & 0.1 		\\
$\gamma$-Ray intensity & 0.1 & 3.1 	 	 & 3.1 		 & 0.1 & 1.5/1.4					 & 3.4       & 0.3 & 3.2   			   			\\
$\gamma$-Ray self-absorption & --  	 	 & \multicolumn{7}{c}{0.2}   	 				   	   	 	 						\\
Summing corrections					& --		& 0.4	& -- & -- & -- & -- & 0.4 & -- \\
Counting statistics          & 0.1 - 1.0 & 0.3 - 0.6 & 0.4 - 0.8 & 2.0-2.6 & 0.4 - 1.4 & 2.6 - 3.1 & 1.2 - 1.4 & 1.2 - 2.4\\
		 					 & 	  	 	 & 	  	     & 	  		 & 	  	   	 & 	         & 	  		 & 	  		 & 	  		\\
Total uncertainty$^c$ 		 &	         & 4.8  	 & 4.8 & 4.1 - 4.5 & 3.7 - 4.0 & 5.4 - 5.8 & 3.6 - 3.9 & 7.3 - 7.6\\
\end{tabular}
\end{ruledtabular}
$^a$ $\gamma$-Ray lines at 216 and 373~keV. \\
$^b$ Not included in final uncertainty, see text.\\
$^c$ Including the respective uncertainty of gold (2.0-2.2\%).
\end{table*}

Significant uncertainties were contributed by the sample position, 
detector efficiencies, and $\gamma$-ray intensities. In the 
activations the position of the samples relative to the Au 
foils was estimated to $\pm$0.25~mm, leading to a 2\% uncertainty 
in the neutron flux. The same uncertainty had to be assigned to 
the efficiency calibration of both detector systems. The largest 
uncertainties were in most cases introduced by the adopted
$\gamma$-ray intensities, $I_\gamma$, an aspect that can be 
improved if more accurate spectroscopic data become available.  

Minor uncertainties arise from the sample masses, which were 
determined to $\pm$0.1~mg, self absorption corrections, and 
counting statistics. The isotopic compositions also exhibit 
rather small uncertainties, except for $^{120}$Te and $^{156}$Dy,
where 11.1\% and 5.4\% are assigned in Ref. \cite{iupac03}, 
respectively (Table~\ref{iso}). Similar to other rare isotopes, 
i.e. $^{184}$Os, $^{190}$Pt, or $^{180}$Ta, the uncertainties 
were conservatively treated to "cover the range of probable 
isotope-abundance variations among different materials as 
well as experimental errors"~\cite{iupac03}. As described in Sec.~\ref{Exp} we have therefore for $^{120}$Te calculated a weighted average of two recent measurements \cite{deLae94,Lee95} and a corrected value from an older measurement \cite{SRL78,deLae94} with a standard deviation of 0.9\%. 

Summing corrections are another potential source of uncertainties. In the measurements with the single HPGe detector these corrections are small due to the low efficiency and can be completely avoided by selecting cascades with only one strong transition. In case of the measurements with the Clover system, the summing corrections were determined by detailed GEANT simulations \cite{Ago03} of the complete setup \cite{clover}. Since these corrections turned out to be rather small (Table IV) the related uncertainties have almost no impact on the final results.

In all cases, the total uncertainties include the 2.0-2.2\% uncertainty of the gold measurements.

\section{Results}\label{Res}

In this section, the results of the present work are presented
in tabular form with a short discussion of the various 
measurements. The comparison with previous results will be
given in the Sec.~\ref{macs} for the Maxwellian average cross 
sections $\langle \sigma \rangle_{kT}$ calculated on the basis of the measured $\langle \sigma_{exp} \rangle$
reported here.

\subsection{$^{102}$Pd(n,$\gamma$)$^{103}$Pd}

Due to the weak $\gamma$ transitions the activated samples were
counted with the Clover system. The $\gamma$ spectra of the Pd
samples were analyzed via the "strongest" transition in $^{103}$Rh
at 357~keV. The second strongest transition at 497~keV was already
too weak and could not be analyzed. The experimental neutron
capture cross section is $\langle \sigma_{exp} \rangle = 376\pm17$ 
mbarn (Table~\ref{tab:xs-pd}).

\begin{table}[!htb]
\caption{\label{tab:xs-pd}Measured (n,$\gamma$) cross sections
of $^{102}$Pd and total uncertainties.}
\renewcommand{\arraystretch}{1.1} 
\begin{ruledtabular}
\begin{tabular}{cc}
Sample & $\langle \sigma_{exp} \rangle$ [mbarn]\\
 & $E_\gamma$= 357 keV \\
\hline
pd-1 & 374 $\pm$ 18 \\
pd-2 & 357 $\pm$ 17 \\
pd-3 & 403 $\pm$ 19 \\
     &              \\
Weighted average & 376 $\pm$ 17 \\
\end{tabular}
\end{ruledtabular}
\end{table}

\subsection{$^{120}$Te(n,$\gamma$)$^{121}$Te}

The Te samples were analyzed via the 573 keV $\gamma$-line from
the $\beta^+$ decay of $^{121}$Te$^{\rm g}$ into $^{121}$Sb. The 
partial cross section to the isomeric state could be determined
only after a waiting time of 80~d, when the large Compton
background around 210 keV, which was observed immediately after 
the irradiation, was sufficiently reduced to reveal the expected 
212 keV line from the IT decay to the ground state. The results 
are 470.6$\pm$22.7~mbarn for the neutron capture cross section to the 
ground-state and 67.6$\pm$2.9~mbarn for the partial cross section 
to the isomeric state, leading to a total ($n, \gamma$) cross 
section of $\langle \sigma_{exp} \rangle$ = 538.2$\pm$25.6 mbarn 
(Table~\ref{tab:xs-te}).

\begin{table}[!htb]
\caption{\label{tab:xs-te}Measured (n,$\gamma$) cross sections 
   of $^{120}$Te and total uncertainties.}
\renewcommand{\arraystretch}{1.1} 
\begin{ruledtabular}
\begin{tabular}{ccc}
Sample & \multicolumn{2}{c}{$\langle \sigma_{exp} \rangle$ [mbarn]}\\
           & $\rightarrow$$^{121}$Te$^{\rm g}$ & $\rightarrow$$^{121}$Te$^{\rm m}$ \\
 & $E_\gamma$= 573 keV & 212 keV \\
\hline
te-1 & 474.4$\pm$22.9  &  66.6$\pm$2.8 \\
te-2 & 465.5$\pm$22.4  &  70.1$\pm$2.9 \\
te-3 & 484.0$\pm$23.2  &  66.5$\pm$2.8 \\
te-4 & 462.2$\pm$22.3  &  65.7$\pm$2.9 \\
te-5 & 468.2$\pm$22.7  &  69.4$\pm$3.1 \\
     &                   &                 \\
Weighted average & 470.6$\pm$22.7 & 67.6$\pm$2.9 \\
Total (n,$\gamma$) cross section & \multicolumn{2}{c}{538.2$\pm$25.6} \\
\end{tabular}
\end{ruledtabular}
\end{table}

\subsection{$^{130}$Ba(n,$\gamma$)$^{131}$Ba}

The $^{130}$Ba cross section has been 
measured via the transitions at 216 and 373~keV from
the $\beta^+$ decay into $^{131}$Cs. The two strongest transitions, 124~keV ($I_\gamma$=29.8 (3)~\%) and 496~keV ($I_\gamma$=48.0 (4)~\%), were not used in this analysis because these lines are affected by coincidence summing effects (summing-out into the 620~keV transition).
Due to the short isomeric half-life of 14.6~min the partial cross section to the isomer 
could not be determined. The resulting total experimental 
cross section is $\langle \sigma_{exp} \rangle$ = 736$\pm$29~mbarn 
(Table~\ref{tab:xs-ba130}).

\begin{table}[!htb]
\caption{\label{tab:xs-ba130}Measured (n,$\gamma$) cross sections of $^{130}$Ba and total uncertainties.}
\renewcommand{\arraystretch}{1.1} 
\begin{ruledtabular}
\begin{tabular}{ccccc}
Sample & \multicolumn{2}{c}{$\langle \sigma_{exp} \rangle$ [mbarn]}\\
 					 & $E_\gamma$= 216 keV      & 373 keV    \\
\hline
ba-1       &  724 $\pm$ 29  &  737 $\pm$ 29 \\
ba-2       &  748 $\pm$ 29  &  752 $\pm$ 30 \\
ba-3       &  718 $\pm$ 27  &  743 $\pm$ 29 \\
		   		 & 	 	   	    	   		 	&  	   		 	\\
Weighted average & \multicolumn{2}{c}{736 $\pm$ 29}							\\
\end{tabular}
\end{ruledtabular}
\end{table}

\subsection{$^{132}$Ba(n,$\gamma$)$^{133}$Ba}
For $^{132}$Ba the partial cross section to the isomeric state 
($t_{1/2}=38.9$~h) could be measured in addition to the total 
(n,$\gamma$) cross section. The latter measurement was 
performed with the Clover detector because of the long half 
life of $^{133}$Ba$^{\rm g}$ ($t_{1/2}=10.52$~yr). The partial 
cross section to the isomer $^{133}$Ba$^{\rm m}$ was
measured via the 276 keV line (99.99\% IT) to  
$\langle \sigma_{exp} \rangle ^{\rm m} = 35.5 \pm 2.0$ mbarn. 
The EC part of the isomeric decay is only 0.0096\% and was, 
therefore, neglected. The total capture cross section of 
$\langle \sigma_{exp} \rangle = 393 \pm 15$ mbarn was determined 
via the strongest transition in the EC decay to $^{133}$Cs 
at 356.0 keV (Table~\ref{tab:xs-ba132}).

\begin{table}[!htb]
\caption{\label{tab:xs-ba132}Measured (n,$\gamma$) cross 
   sections of $^{132}$Ba and total uncertainties.}
\renewcommand{\arraystretch}{1.1} 
\begin{ruledtabular}
\begin{tabular}{ccc}
Sample & \multicolumn{2}{c}{$\langle \sigma_{exp} \rangle$ [mbarn]}\\
     & $\rightarrow$$^{133}$Ba$^{\rm tot}$ & $\rightarrow$$^{133}$Ba$^{\rm m}$ \\
 & $E_\gamma$= 356 keV     & 276 keV \\
\hline
ba-1 & 396.0$\pm$15.2  &  32.7$\pm$1.9 \\
ba-2 & 403.9$\pm$15.6  &  39.6$\pm$2.3 \\
ba-3 & 381.8$\pm$13.7  &  35.4$\pm$1.9 \\
     &                   &                 \\
Weighted average & 392.8$\pm$14.8  &  35.5$\pm$2.0 \\
\end{tabular}
\end{ruledtabular}
\end{table}

\subsection{$^{156}$Dy(n,$\gamma$)$^{157}$Dy}

The (n,$\gamma$) cross section of $^{156}$Dy was measured via 
the strongest line in the decay of $^{157}$Dy at 326 keV. The 
uncertainty of the measured value $\langle \sigma_{exp} \rangle$ = 1641 $\pm$ 117 mbarn
(Table~\ref{tab:xs-dy}) is dominated by the 
contributions from the $\gamma$-ray intensity (3.2\%) and
from the isotopic abundance (5.4\%). 

\begin{table}[!htb]
\caption{\label{tab:xs-dy}Measured (n,$\gamma$) cross sections 
   of $^{156}$Dy and total uncertainties.}
\renewcommand{\arraystretch}{1.1} 
\begin{ruledtabular}
\begin{tabular}{cc}
Sample & $\langle \sigma_{exp} \rangle$ [mbarn] \\
 & $E_\gamma$= 326 keV \\
\hline
dy-1 & 1669 $\pm$ 121  \\
dy-2 & 1638 $\pm$ 114   \\
dy-3 & 1619 $\pm$ 114   \\
     &                 \\
Weighted average & 1641 $\pm$ 117 \\
\end{tabular}
\end{ruledtabular}
\end{table}

\subsection{Isomeric ratios}

Isomeric ratios 
\begin{eqnarray}
R_\mathrm{iso}=\frac{\langle \sigma_{exp} \rangle ^{\rm m}}{\langle \sigma_{exp} \rangle ^{\rm tot}}
\end{eqnarray}
were calculated for $^{121}$Te and $^{133}$Ba from the measured 
partial and total (n,$\gamma$) cross sections. The present results 
at $kT$=25~keV are 0.126$\pm$0.012 and 0.090$\pm$0.009, respectively. These results are
compatible with the thermal values at $kT$= 25 meV, which are
0.145$\pm$0.026 for $^{121}$Te and 0.071 for $^{133}$Ba \cite{mugh81,mugh06}. 

\section{\label{macs}Maxwellian averaged cross sections}
\subsection{General remarks}

In an astrophysical environment with temperature $T$, interacting 
particles are quickly thermalized by collisions in the stellar 
plasma, and the neutron energy distribution can be described by a Maxwell-Boltzmann spectrum:
\begin{eqnarray}
\Phi = dN/dE_n \sim \sqrt{E_n} \cdot {\rm e}^{-E_n /kT} \label{eq:phi}.
\end{eqnarray}                 
The experimental neutron spectrum of the $^7$Li($p,n$)$^7$Be
reaction simulates the energy dependence of the flux $v \cdot \Phi \sim E_n \cdot e^{-E_n /kT}$ with
$kT$=25.0 $\pm$ 0.5~keV almost perfectly \cite{raty88}. However, 
the cutoff at $E_n$= 106~keV and small deviations from the 
shape of the ideal Maxwellian spectrum require a correction 
of the measured cross section $\langle\sigma_{exp}\rangle$ for obtaining a 
true Maxwellian average, $\langle\sigma\rangle$$_{25~keV}$. This correction
is determined by means of the energy-dependent cross 
sections from data libraries.


\subsection{Evaluated cross sections from data libraries}
\label{sec:libraries}

The corrections for the spectrum differences as well as the 
extrapolations of the MACS to lower and higher values of $kT$
were determined with the evaluated energy-dependent cross 
sections, $\sigma$(E$_n$), from the data libraries provided 
by the online database JANIS 3.0 ("Java-based Nuclear Information Software", www.nea.fr/janis/) \cite{janis30}. The libraries used were the "Joint Evaluated Fission and Fusion 
General Purpose File" ({\sc JEFF~3.0A} and {\sc JEFF~3.1}, 
www.nea.fr/html/dbdata/{\sc JEFF/}), the "Japanese Evaluated 
Nuclear Data Library" ({\sc JENDL~3.3} \cite{jendl33},
wwwndc.tokai-sc.jaea.go.jp/jendl/), and the "Evaluated 
Nuclear Data File" ({\sc ENDF-B/VII.0} \cite{endfb7}, 
www.nndc.bnl.gov/), which are partially based on experimental 
resonance parameters.

\begin{table}[!htb]
\caption{Upper limits (in eV) of the resolved resonance region 
in different databases. Cases without resonance information are indicated by empty entries.} \label{tab:RRR}
\renewcommand{\arraystretch}{1.1} 
\begin{ruledtabular}
 \begin{tabular}{ccccc}
Isotope & JEFF 3.0A & JEFF 3.1 & JENDL 3.3 & ENDF-B/VII.0 \\
\hline
$^{102}$Pd & 397 & 397 & 250 & 820 \\
$^{120}$Te & -- & -- & -- & -- \\
$^{130}$Ba & 2030 & 2530 & 2530 & 2800 \\
$^{132}$Ba & -- & -- & -- & 130 \\
$^{156}$Dy & 101.3 & -- & --  & 91 \\
\end{tabular}
\end{ruledtabular}
\end{table}

For the investigated cases, the most recent data for 
$\sigma$(E$_n$) are provided by {\sc ENDF-B/VII.0}, which 
makes use of resonance parameters from the latest evaluation 
\cite{mugh06}. The differences between the four data libraries 
with respect to the resolved resonance region (RRR) are
summarized in Table~\ref{tab:RRR}. Table~\ref{tab:RRR2} shows 
that the contributions of the RRR to the respective Maxwellian 
averaged cross sections for $kT$=5-100~keV are almost negligible, 
except for the lower temperatures in $^{130}$Ba, where the RRR 
reaches up to about 3~keV. In all libraries, the ($n, \gamma$)
cross sections in the unresolved resonance region (URR) were 
obtained by Hauser-Feshbach (HF) calculations. Since this 
region contributes the most important part to the extrapolation 
towards the higher temperatures of the $p$-process, it 
is discussed in more detail.

The Hauser-Feshbach calculations in the JENDL-3.3 
evaluations were performed with the statistical model code 
{\sc CASTHY} \cite{casthy} for the isotopes investigated
in this work. More specifically, the $\gamma$-ray strength 
function for $^{132}$Ba was adjusted to reproduce the available 
experimental capture cross section of Ref. \cite{BPS79}. 
For JEFF-3.0A, the HF calculations are not documented, and 
in JEFF-3.1 an unspecified HF prediction with a Moldauer 
potential was used  for $^{102}$Pd and $^{120}$Te, whereas 
the {\sc CASTHY} code was also used for the URR of 
$^{130,132}$Ba. In ENDF/B-VII.0, the URR in $^{102}$Pd was 
obtained with the {\sc GNASH} code \cite{gnash}, and for 
$^{156}$Dy results from {\sc EMPIRE} \cite{empire} were used. 
For $^{120}$Te and $^{130,132}$Ba the respective URR 
cross sections from JENDL-3.3 have been adopted also in ENDF/B-VII.0.

\begin{table*}[!htb]
\caption{\label{tab:RRR2} Contribution of the resolved resonance 
region to the Maxwellian averaged cross sections (in \%) for $kT$=5-100~keV.}
\renewcommand{\arraystretch}{1.1} 
\begin{ruledtabular}
\begin{tabular}{lccccccccccc}
$kT$ [keV]	 & 5     & 10  & 15  & 20  & 25  & 30  & 40  & 50  & 60  & 80  & 100 \\
\hline 
 			 &  \multicolumn{11}{c}{$^{102}$Pd}		  		  		   			 \\ 	
JEFF 3.0A    & 2.5   & 0.9 & 0.5 & 0.3 & 0.3 & 0.2 & 0.1 & 0.1 & 0.1 & 0   & 0   \\
JEFF 3.1     & 2.5   & 0.9 & 0.5 & 0.3 & 0.3 & 0.2 & 0.1 & 0.1 & 0.1 & 0   & 0   \\
JENDL 3.3    & 1.4   & 0.5 & 0.3 & 0.2 & 0.1 & 0.1 & 0.1 & 0   & 0   & 0   & 0   \\ 
ENDF-B/VII.0 & 4.4   & 1.8 & 1.0 & 0.7 & 0.5 & 0.4 & 0.3 & 0.2 & 0.1 & 0.1 & 0.1 \\  					  					  					  					 \\
 						&  \multicolumn{11}{c}{$^{130}$Ba}    		  			 \\ 
JEFF 3.0A    & 8.1   & 3.8 & 2.4 & 1.6 & 1.2 & 1.0 & 0.6 & 0.5 & 0.4 & 0.2 & 0.2 \\
JEFF 3.1     & 12.5  & 5.7 & 3.2 & 2.0 & 1.5 & 1.2 & 0.9 & 0.7 & 0.5 & 0.3 & 0.2 \\
JENDL 3.3    & 12.5  & 5.7 & 3.2 & 2.0 & 1.5 & 1.2 & 0.9 & 0.7 & 0.5 & 0.3 & 0.2 \\
ENDF-B/VII.0 & 11.6  & 5.2 & 3.1 & 2.1 & 1.5 & 1.1 & 0.7 & 0.5 & 0.4 & 0.2 & 0.2 \\  					  					  					  					 \\
 			 &  \multicolumn{11}{c}{$^{132}$Ba}  	  				  		  	 \\  	
ENDF-B/VII.0 & 1.5   & 0.5 & 0.3 & 0.2 & 0.1 & 0.1 & 0.1 & 0   & 0   &  0  & 0   \\  					  					  					  					 \\
 			 &  \multicolumn{11}{c}{$^{156}$Dy}  	  				   			 \\ 
JEFF 3.0A    & 0.7   & 0.3 & 0.2 & 0.1 & 0.1 & 0.1 & 0   & 0   & 0   & 0   & 0   \\
ENDF-B/VII.0 & 0.5   & 0.2 & 0.1 & 0.1 & 0.1 & 0   & 0   & 0   & 0   & 0   & 0   \\  	
\end{tabular}
\end{ruledtabular}
\end{table*}

\subsection{Calculation of Maxwellian average cross sections}

In a first step, the evaluated cross sections were folded with 
the experimental neutron spectrum. The ratios of the measured cross 
sections and the corresponding averages of the evaluated data,
\begin{equation}
F_\mathrm{norm}=\frac{\langle \sigma_\mathrm{exp} \rangle}{\langle \sigma_\mathrm{eval}\rangle} \quad,
\end{equation}
are listed in Table \ref{tab:NF} for all investigated cases. 
Since the RRR contributes an almost negligible part to the 
MACS at higher temperatures, $F_\mathrm{norm}$ can 
be used in very good approximation as normalization 
factor for the URR. This holds even for $^{130}$Ba, where 
the RRR contributes sensibly at lower thermal energies, 
because in this case the $F_\mathrm{norm}$ values are very close to unity.

\begin{table}
\caption{Normalization factors 
$F_\mathrm{norm} = \langle \sigma_\mathrm{exp}\rangle / \langle \sigma_\mathrm{eval}\rangle$ 
for adjusting the evaluated cross sections in the URR. The 
factors $\langle \sigma_\mathrm{exp} \rangle / \langle \sigma_\mathrm{HF} \rangle$ for the pure Hauser-Feshbach models {\sc MOST} 
\cite{most05} and {\sc NON-SMOKER} \cite{rath00,nons} 
are also listed. }
\label{tab:NF}
\renewcommand{\arraystretch}{1.1} 
\begin{ruledtabular}
\begin{tabular}{lccccc}
 				   &$^{102}$Pd &$^{120}$Te &$^{130}$Ba &$^{132}$Ba & $^{156}$Dy \\
\hline 
{\sc JEFF 3.0A}    & 1.852	   & --	   & 0.990	   & 0.870 	   & 1.069	\\
{\sc JEFF 3.1}     & 1.852	   & 1.263	   & 1.024	   & 0.869 	   & --		\\
{\sc JENDL 3.3}    & 0.967	   & 1.843	   & 1.025	   & 0.873 	   & --		\\
{\sc ENDF-B/VII.0} & 0.818	   & 1.843	   & 1.028	   & 0.873 	   & 1.059	\\
	  			   & 		   & 		   & 		   & 		   & 		\\
{\sc NON-SMOKER}   & 0.985	   & 0.978	   & 1.019	   & 0.840 	   & 1.518 	\\
{\sc MOST 2005}$^a$ & 0.550	   & 1.752	   & 1.505	   & 1.746 	   & 0.781  \\
\end{tabular}
\end{ruledtabular}
$^a$Including stellar enhancement factors without further specification.
\end{table}

In principle, a full normalization of the evaluated data would 
change the thermal cross section as well as the resolved 
resonances. Since these data are (partially) based on experimental 
information, the contribution from the RRR has been decoupled 
from the normalization procedure and the MACS were calculated 
using the RRR contributions listed in Table \ref{tab:RRR2}. 
The contribution from the URR was then determined 
from the renormalized part of the evaluated cross sections,
which consist of theoretical data obtained in HF calculations:
\begin{equation}
\langle\sigma\rangle_{kT} = \langle\sigma\rangle_{kT}^{\rm RRR}
                       + F_\mathrm{norm} \cdot \langle\sigma\rangle_{kT}^{\rm URR}. \label{eq:total}
\end{equation}
This expression is equivalent to obtaining a Maxwellian average
from the energy-differential cross section $\sigma(E_\mathrm{n})$ after 
only the URR has been modified (Eq.~\ref{eq:macs}). The respective 
Maxwellian averaged cross sections are
\begin{eqnarray}
\langle\sigma\rangle_{kT}=\frac{2}{\sqrt{\pi}} \left[\frac{\int^\mathrm{RRR}
\sigma(E_\mathrm{n}) \cdot E_\mathrm{n} \cdot e^{-E_\mathrm{n}/(kT)}\,dE_\mathrm{n}}{\int^\mathrm{RRR+URR}
E_\mathrm{n} \cdot e^{-E_\mathrm{n}/(kT)}\,dE_\mathrm{n}}~+ \right. \nonumber \\
\left. \frac{\int^\mathrm{URR} F_\mathrm{norm} \cdot \sigma(E_\mathrm{n}) \cdot E_\mathrm{n} \cdot e^{-E_\mathrm{n}/(kT)}\,dE_\mathrm{n}}{\int^\mathrm{RRR+URR}
E_\mathrm{n} \cdot e^{-E_\mathrm{n}/(kT)}\,dE_\mathrm{n}} \right] \quad. \label{eq:macs}
\end{eqnarray}

The values for thermal energies between $kT$=5 and 100 keV in Table~\ref{tab:macs} were derived by normalization with the respective factors $F_{norm}$ listed in Table \ref{tab:NF}. Evaluations yielding the same normalization factors are (obviously) based on the same resonance parameters (see discussion in Sec.~\ref{sec:libraries}). Additionally the original and normalized values from the recommendations in Bao et al. \cite{bao00} are listed in Table~\ref{tab:macs} for comparison. Note that the previous semi-empirical estimates for $^{102}$Pd, $^{120}$Te, and $^{132}$Ba in the Bao et al. compilation \cite{bao00} were based on scaled NON-SMOKER predictions (see Sec.~\ref{sec:comp} for further details).

\begin{table*}
\caption{\label{tab:macs} Maxwellian averaged cross sections $\langle \sigma \rangle_{kT}$ (in mbarn) and stellar enhancement factors (f$^*$) for $kT$=5-100~keV. Listed are the normalized Maxwellian averaged cross sections from the evaluations, and the original and normalized values from Ref.~\cite{bao00}.}
\renewcommand{\arraystretch}{1.2} 
\begin{ruledtabular}
\begin{tabular}{lccccccccccc}
$kT$ [keV]	 & 5     & 10    & 15     & 20 & 25    & 30  & 40  & 50 & 60   & 80   & 100    	 \\
\hline 
 			 &  \multicolumn{11}{c}{$^{102}$Pd}  	  				  		  		  		  		   			 \\ 	
Bao \cite{bao00} & 894 & 657 & 540 & 466 & 414 & 375$\pm$118$^a$ & 320 & 283 & 257 & 222 & 199 \\
Bao norm. & 875 & 643 & 529 & 456 & 405 & 367$\pm$17 & 313 & 277 & 252 & 217 & 195 \\
ENDF/B-VII.0 			& 835$\pm$41 & 611$\pm$29 & 511$\pm$24 & 449$\pm$21 & 405$\pm$18 & 371$\pm$17 & 323$\pm$15 & 292$\pm$13 & 270$\pm$12 & 243$\pm$11 & 225$\pm$10 \\
JEFF-3.0A/3.1    	& 913$\pm$41 & 657$\pm$30 & 534$\pm$24 & 458$\pm$21 & 404$\pm$18 & 365$\pm$16 & 311$\pm$14 & 275$\pm$12 & 250$\pm$11 & 218$\pm$10 & 198$\pm$9 \\
JENDL-3.3   			& 905$\pm$36 & 644$\pm$27 & 527$\pm$23 & 455$\pm$20 & 406$\pm$18 & 370$\pm$17 & 321$\pm$15 & 289$\pm$13 & 267$\pm$12 & 239$\pm$11 & 222$\pm$10 \\ 
f$^*$ \cite{rath00}   & 1.000 & 1.000 &  1.000 & 1.000 & 1.000 & 1.000 & 1.000 & 1.000 & 1.000 & 1.003 & 1.011 	 \\
 			 &       &       &        &       &       &               &       &       &       &       &       	 \\ 	
 			 &  \multicolumn{11}{c}{$^{120}$Te}  	  				  		  		  		  		  			 \\ 
Bao \cite{bao00} 	& 1037 & 708 & 578 & 504 & 455 & 420$\pm$103$^a$ & 372   & 341   & 318   & 286   & 263 	 \\
Bao norm. 				& 1331 & 909 & 742 & 647 & 584 & 539$\pm$26 & 477 & 438 & 408 & 367 & 337 \\
ENDF/B-VII.0 			& 1319$\pm$58 & 919$\pm$42 & 749$\pm$35 & 649$\pm$31 & 583$\pm$28 & 535$\pm$26 & 472$\pm$22 & 431$\pm$21 & 403$\pm$19 & 368$\pm$18 & 348$\pm$17 \\ 
JEFF-3.1    			& 1215$\pm$63 & 880$\pm$44 & 733$\pm$36 & 645$\pm$31 & 585$\pm$28 & 540$\pm$26 & 478$\pm$22 & 436$\pm$21 & 407$\pm$19 & 368$\pm$18 & 345$\pm$17 \\
JENDL-3.3   			& 1319$\pm$58 & 919$\pm$42 & 749$\pm$35 & 649$\pm$31 & 583$\pm$28 & 535$\pm$26 & 472$\pm$22 & 431$\pm$21 & 403$\pm$19 & 368$\pm$18 & 348$\pm$17 \\ 
f$^*$ \cite{rath00}    & 1.000 & 1.000 &  1.000 & 1.000 & 1.000 & 1.000 & 1.000 & 1.000 & 1.000 & 1.003 & 1.010 	 \\
 			 &       &       &        &       &       &               &       &       &       &       &       	 \\ 	
 						&  \multicolumn{11}{c}{$^{130}$Ba}  		  		  		  		  		  			 \\ 
Bao \cite{bao00} 	& 2379 & 1284 & 1031 & 901 & 818 & 760$\pm$110 & 683   & 634   & 601   & 556   & 526 	 \\
Bao norm. 				& 2333 & 1259 & 1011 & 884 & 802 & 745$\pm$29 & 670 & 622 & 589 & 545 & 516 \\
ENDF/B-VII.0 			& 1600$\pm$67 & 1163$\pm$45 & 979$\pm$38  & 874$\pm$34 & 805$\pm$31 & 756$\pm$29 & 687$\pm$27 & 642$\pm$25 & 610$\pm$24 & 568$\pm$22 & 542$\pm$21 \\ 
JEFF-3.0/A    		& 1862$\pm$64 & 1291$\pm$45 & 1040$\pm$38 & 894$\pm$34 & 797$\pm$31 & 728$\pm$29 & 635$\pm$27 & 575$\pm$25 & 532$\pm$24 & 470$\pm$22 & 425$\pm$21 \\
JEFF-3.1 					& 1634$\pm$69  & 1171$\pm$45 & 982$\pm$38 & 874$\pm$34 & 804$\pm$31 & 754$\pm$29 & 686$\pm$27 & 640$\pm$25 & 608$\pm$24 & 566$\pm$22 & 540$\pm$21 \\ 
JENDL-3.3 				& 1634$\pm$69  & 1171$\pm$45 & 982$\pm$38 & 874$\pm$34 & 804$\pm$31 & 754$\pm$29 & 686$\pm$27 & 640$\pm$25 & 608$\pm$24 & 566$\pm$22 & 540$\pm$21 \\
f$^*$ \cite{rath00}    & 1.000 & 1.000 &  1.000 & 1.000 & 1.000 & 1.000 & 1.000 & 1.002 & 1.006 & 1.024 & 1.056 	 \\
 			 &       &       &        &       &       &               &       &       &       &       &       	 \\ 	
 			 &  \multicolumn{11}{c}{$^{132}$Ba}  	  				  		  		  		  		  			 \\  	
Bao \cite{bao00} 	& 1029 & 659 & 526 & 455 & 410 & 379$\pm$137$^a$ & 339   & 315   & 298   & 276   & 261 	 \\
Bao norm. 				& 1070 & 685 & 547 & 473 & 426 & 394$\pm$15 & 353 & 328 & 310 & 287 & 271 \\
ENDF/B-VII.0 			&  929$\pm$35 & 641$\pm$24 & 530$\pm$20 & 468$\pm$18 & 427$\pm$16 & 398$\pm$15 & 358$\pm$14 & 331$\pm$13 & 313$\pm$12 & 290$\pm$11 & 277$\pm$11 \\ 
JEFF-3.0A/3.1			&  915$\pm$35 & 637$\pm$24 & 528$\pm$20 & 467$\pm$18 & 426$\pm$16 & 397$\pm$15 & 357$\pm$14 & 331$\pm$13 & 312$\pm$12 & 290$\pm$11 & 276$\pm$11 \\
JENDL-3.3					&  915$\pm$35 & 637$\pm$24 & 528$\pm$20 & 467$\pm$18 & 426$\pm$16 & 397$\pm$15 & 357$\pm$14 & 331$\pm$13 & 312$\pm$12 & 290$\pm$11 & 276$\pm$11 \\
f$^*$ \cite{rath00}   & 1.000 & 1.000 &  1.000 & 1.000 & 1.000 & 1.000 & 1.000 & 1.001 & 1.004 & 1.018 & 1.040 	 \\
 			 &       &       &        &       &       &               &       &       &       &       &       	 \\ 	
 			 &  \multicolumn{11}{c}{$^{156}$Dy}  	  				  		  		  		  		  			 \\ 
Bao \cite{bao00}  & 5442 & 2712 & 2126 & 1850 & 1682 & 1567$\pm$145 & 1412 & 1307  & 1229  & 1117  & 1039 	 \\
Bao norm. 				& 5742 & 2862 & 2243 & 1952 & 1775 & 1653$\pm$118 & 1490 & 1379 & 1297 & 1179 & 1096 \\
ENDF/B-VII.0 			& 4742$\pm$336 & 3012$\pm$214 & 2355$\pm$168 & 2001$\pm$143 & 1775$\pm$126 & 1616$\pm$115 & 1408$\pm$100 & 1274$\pm$91 & 1180$\pm$84 & 1052$\pm$75 & 968$\pm$69 \\ 
JEFF-3.0A					& 4060$\pm$288 & 2710$\pm$193 & 2216$\pm$158 & 1948$\pm$139 & 1775$\pm$126 & 1651$\pm$118 & 1478$\pm$105 & 1359$\pm$97 & 1269$\pm$91 & 1142$\pm$81 & 1056$\pm$75 \\
f$^*$ \cite{rath00}  	 & 1.000 & 1.000 &  1.001 & 1.007 & 1.022 & 1.046 & 1.106 & 1.163 & 1.210 & 1.283 & 1.341	 \\
\end{tabular}
$^a$ Semi-empirical estimates.
\end{ruledtabular}
\end{table*}

As can be seen in Table~\ref{tab:macs} extrapolation to lower or higher energies reveal large differences. For this reason we cannot recommend one or the other evaluation in this paper and leave it to the reader which energy dependence to use. 

The uncertainties given for the evaluations were derived from Eq.~\ref{eq:total}. For the contribution of the RRR we assumed a conservative uncertainty of 20\%. The uncertainty of each data point in the evaluation is not provided, so $\Delta\langle\sigma_{eval}\rangle$ must be set to 0 and $\Delta F_{norm}$ is defined as $\frac{\Delta\langle\sigma_{exp}\rangle}{\langle\sigma_{eval}\rangle}$. $\Delta \langle\sigma\rangle_{kT}^{\rm URR}$ is the uncertainty of the URR which is based on Hauser-Feshbach predictions. The estimate for this uncertainty is even more difficult, but could be done in principle by comparing different Hauser-Feshbach models which use different input parameters, as in our cases (see discussion in Sec.~\ref{sec:libraries}). The large deviations among the different evaluations towards higher energies reflect the influence of the input parameters, but for the individual evaluations we set $\Delta \langle\sigma\rangle_{kT}^{\rm URR}$=0. Thus the quoted total uncertainty $\Delta$$\langle\sigma\rangle_{kT}$ is
\begin{equation}
\Delta\langle\sigma\rangle_{kT} = \sqrt{\left(\Delta\langle\sigma\rangle_{kT}^{\rm RRR}\right)^2 + \left(\Delta F_{norm} \cdot \langle\sigma\rangle_{kT}^{\rm URR} \right)^2 } \label{eq:deltamacs} .
\end{equation}
In view of the remaining uncertainties, in particular at higher neutron energies, time-of-flight (TOF) data with experimental uncertainties are needed to replace the present extrapolations. 

Since experimentally determined cross sections refer only 
to target nuclei in their ground states, the effective 
stellar cross sections have to be corrected for the fact 
that low-lying excited nuclear states can be thermally 
populated in the hot stellar photon bath. This is achieved 
by introducing the stellar enhancement factor (Eq.~\ref{eq:sef})
\begin{equation}
f^{*}(T) = \frac{\langle \sigma \rangle ^{*}}{\langle \sigma \rangle ^\mathrm{lab}}=
\frac{\sigma^*}{\sigma^\mathrm{lab}} \quad, \label{eq:sef}
\end{equation}
where the stellar cross section
\begin{equation}
\sigma^{*}=\frac{\sum_{\mu} (2J_\mu +1) e^{-E_\mu /(kT)}
\sum_{\nu} \sigma^{\mu\nu}}{\sum_\mu (2J_\mu +1) e^{-E_\mu /(kT)}}
\end{equation}
accounts for the transitions of thermally populated target states 
$\mu$ to all possible final states $\nu$, whereas the laboratory cross
section $\sigma ^\mathrm{lab}=\sum_\nu \sigma^{0\nu}$ only includes the
ground state of the target nuclei.
The stellar enhancement factors are tabulated, e.g., in Refs.\
\cite{rath00,nons,bao00}, and can increase strongly 
with temperature. While the values remain close to unity under typical $s$-process conditions, they 
become significantly larger at the higher 
temperatures of the $p$-process (Table~\ref{tab:sef-p}).

\subsection{Comparison of 30 keV MACS with previous data}\label{sec:comp}

The comparison with previous recommendations \cite{bao00,alle71}, 
experimental data \cite{BPS79,Bee85}, and theoretical predictions 
\cite{holm76,harr81,zhao88,rath00,most05} is summarized in 
Table~\ref{tab:comp} for the MACS at $kT=30$ keV. 

\begin{table*}[!htb]
\caption{Comparison of Maxwellian averaged cross sections 
(in mbarn) at $kT$=30~keV. }\label{tab:comp}
\renewcommand{\arraystretch}{1.2} 
\begin{ruledtabular}
\begin{tabular}{lccccc}
Reference  	    & $^{102}$Pd       & $^{120}$Te       & $^{130}$Ba       & $^{132}$Ba       & $^{156}$Dy 	      \\
\hline
Bao norm.				& 367$\pm$17 & 539$\pm$26 & 745$\pm$29 & 394$\pm$15 & 1653$\pm$118 \\
ENDF/B-VII.0		& 371$\pm$17 & 535$\pm$26 & 756$\pm$29 & 398$\pm$15 & 1616$\pm$115 \\ 
JEFF-3.0A				& 365$\pm$16 & --					& 728$\pm$29 & 397$\pm$15 & 1651$\pm$118 \\ 
JEFF-3.1				& 365$\pm$16 & 540$\pm$26 & 754$\pm$29 & 397$\pm$15 & -- \\ 
JENDL-3.3				& 370$\pm$17 & 535$\pm$26 & 754$\pm$29 & 397$\pm$15 & -- \\ 
						  		 & 			   		&                  &                  &                  &        	    	  \\		
	\multicolumn{6}{c}{Experimental data}				 					  \\
Bradley {\it et al.} \cite{BPS79}& 	--		        &	--	           & 	761$\pm$62 	  &	--		 		 &		--			  \\
Beer \cite{Bee85}  	 			 &	--		        &	--	     	   &	--       	  & 	--		     & 1589$\pm$145 	  \\
						  		 & 			        &             	   &           		  &                  &     	    		  \\				  
  \multicolumn{6}{c}{Recommended data in compilations}	 					 					  \\					
Bao {\it et al.} \cite{bao00}  & 375$\pm$118$^a$  & 420$\pm$103$^a$  & 760$\pm$110 	& 379$\pm$137$^a$  & 1567$\pm$145  \\ 
Allen \cite{alle71} 			 & 320              & 400         	   & 2000        	  & 650              & 870 				  \\
						  		 & 			   		&	         	   &             	  &                  &        	    	  \\				  				\multicolumn{6}{c}{Theoretical predictions}			 				\\					
Holmes {\it et al.} \cite{holm76}& 247 				& 275 			   & 397        	  & 250         	 & 1840         	  \\
Harris \cite{harr81} 	         & 363   			& 776 			   & 1012       	  & 442         	 & 1637				  \\
Zhao {\it et al.} \cite{zhao88}  & 137$\pm$45  		& 293$\pm$96	   & -- 			  & 280$\pm$92 		 & 850$\pm$280  	  \\
{\sc NON-SMOKER} \cite{rath00}   & 323   			& 506 			   & 605         	  & 374         	 & 1190				  \\
{\sc MOST} \cite{most05}$^b$ 	 & 665				& 307 			   & 490         	  & 227         	 & 2126				  \\
\end{tabular}
\end{ruledtabular} 
$^a$ Semi-empirical estimates.\\
$^b$ Includes unspecified stellar enhancement factors.
\end{table*}

The cross sections of $^{102}$Pd, $^{120}$Te, and $^{132}$Ba  
had not been measured in the stellar energy range so far. 
Therefore, the recommended cross sections in the compilation 
of Bao {\it et al.} \cite{bao00} are semi-empirical estimates, 
using {\sc NON-SMOKER} results \cite{rath00} normalized to 
the local cross section systematics of neighboring nuclei. 

The only previous experimental value for $^{130}$Ba in the keV region 
was measured with a filtered neutron beam of $24 \pm 2$ keV 
\cite{BPS79}. The result of $715\pm58$ mbarn was transformed 
into a MACS at $kT=30$ keV of 761 mbarn \cite{bao00}, in good 
agreement with the more accurate value of this work.
 
Also for $^{156}$Dy a previous measurement has been reported 
\cite{Bee85} that was performed with the same activation 
technique used here, but only a single activation had 
been made and the result was given as a preliminary value. 
Nevertheless, there is fair agreement with the result 
obtained in the present series of activations. 

The measurements cover a considerable mass range and consider nuclei
with different properties which makes a comparison to predictions
interesting but difficult to interpret.
The {\sc NON-SMOKER} predictions for $^{102}$Pd, $^{120}$Te, 
$^{130}$Ba, and -- to some extent -- for $^{132}$Ba have been 
confirmed by the experimental results, but the prediction for 
$^{156}$Dy is too low by 28\%, similar to the situation for 
$^{160}$Dy. These isotopes are strongly deformed but so are the 
heavier, stable Dy isotopes. Since the level density of the proton-rich isotopes is high one might expect that the Hauser-Feshbach model would be more reliable. Accordingly, the HF parameterization has to be checked in these cases.
The contribution of single resonances, implicitly included in the measurement of the MACS, may be stronger than
predicted by the Hauser-Feshbach approach.

The predictions from {\sc MOST} had to be derived from the stellar reaction 
rates given in Ref.~\cite{most05}, which already include unspecified stellar enhancement 
factors (see Eq.~\ref{eq:sef}). These factors should be close to unity for $kT$=30 keV.
The values from {\sc MOST} \cite{most05} show significant deviations 
from the measured data for all considered nuclei, ranging from as much as -45\% for 
$^{102}$Pd to +75\% for $^{132}$Ba at $kT=30$ keV.

\subsection{Extrapolation to $p$-process energies}
Maxwellian averaged cross sections are also needed at
the higher temperatures of the $p$-process of $2-3$ GK, 
corresponding to thermal energies of $kT=170-260$ keV.
Using the energy-dependencies of the normalized evaluated 
cross sections discussed before, this extrapolation 
yields the MACS listed in Table~\ref{tab:sef-p}. In this 
energy range any contributions from the RRR are 
completely negligible but the uncertainties introduced
by the extrapolation become significant.

\begin{table}
\caption{\label{tab:sef-p}Maxwellian averaged cross 
  sections $\langle \sigma \rangle_{kT}$ (in mbarn) and stellar enhancement factors (f$^*$)
  \cite{rath00,nons} at $p$-process temperatures.}
\begin{ruledtabular}
\begin{tabular}{lccc}
$kT$ [keV] & 170               & 215 			   & 260   	  		   \\
\hline
$^{102}$Pd & & & \\	   	 				
ENDF/B-VII.0 			& 191$\pm$9 & 177$\pm$8 & 168$\pm$8 \\
JEFF-3.0A/3.1    	& 164$\pm$7 & 151$\pm$7 & 141$\pm$6 \\
JENDL-3.3   			& 192$\pm$9 & 182$\pm$8 & 173$\pm$8 \\ 
f$^*$ \cite{rath00}  	   & 1.11 	           & 1.19 	 		   & 1.27 	  		   \\
           & 			       & 		 		   &			  	   \\
 $^{120}$Te & & &  \\  	   					
ENDF/B-VII.0 / JENDL-3.3	& 311$\pm$15 & 296$\pm$14 & 284$\pm$14 \\
JEFF-3.1    							& 304$\pm$15 & 291$\pm$14 & 281$\pm$14 \\
f$^*$ \cite{rath00}	   & 1.10 	           & 1.18 	 		   & 1.25 	  		   \\
           & 			       & 		 		   &			  	   \\
 $^{130}$Ba & & & \\
ENDF/B-VII.0 			 & 505$\pm$20 & 500$\pm$20 & 503$\pm$20 \\
JEFF-3.0A    			 & 323$\pm$20 & 280$\pm$20 & 248$\pm$20 \\
JEFF-3.1/JENDL-3.3 & 504$\pm$20 & 499$\pm$20 & 501$\pm$20 \\
f$^*$ \cite{rath00}  	   & 1.23 	           & 1.35 	 		   & 1.42 	  		   \\
           & 			       & 		 		   &			  	   \\
 $^{132}$Ba & & & \\	   					 
Evaluations 			& 258$\pm$10 & 254$\pm$10 & 253$\pm$10 \\
f$^*$ \cite{rath00}   	   & 1.16 	   		   & 1.23 	 		   & 1.28 	  		   \\
           & 			   	   & 		 		   &			  	   \\
$^{156}$Dy  & & & \\
ENDF/B-VII.0 			& 819$\pm$58 & 772$\pm$55 & 738$\pm$53 \\
JEFF-3.0A    			& 895$\pm$64 & 841$\pm$60 & 802$\pm$57 \\
f$^*$ \cite{rath00}      & 1.50 	   		   & 1.55 	 		   & 1.56 	  		   \\
\end{tabular}
\end{ruledtabular}
\end{table}

The stellar reaction rate can be determined via
\begin{eqnarray}
N_A \cdot \langle\sigma v\rangle= 26445.5 \cdot f^{*} \cdot \langle\sigma\rangle_{kT}
\cdot \sqrt{kT/\mu} \quad, \label{eq:rate4}
\end{eqnarray}
with $\mu$ being the reduced mass, $f^{*}$ the stellar enhancement factor (Eq.~\ref{eq:sef}), and $N_A$ the Avogadro constant.
With the given numerical prefactor, the units for the MACS, $\langle\sigma\rangle_{kT}$, the thermal 
energy $kT$, and the reaction rate N$_A$$\langle\sigma v\rangle$ are [mbarn], [keV], and [cm$^3$~mole$^{-1}$~s$^{-1}$], respectively.

\begin{figure*}[p]
\includegraphics[scale=1]{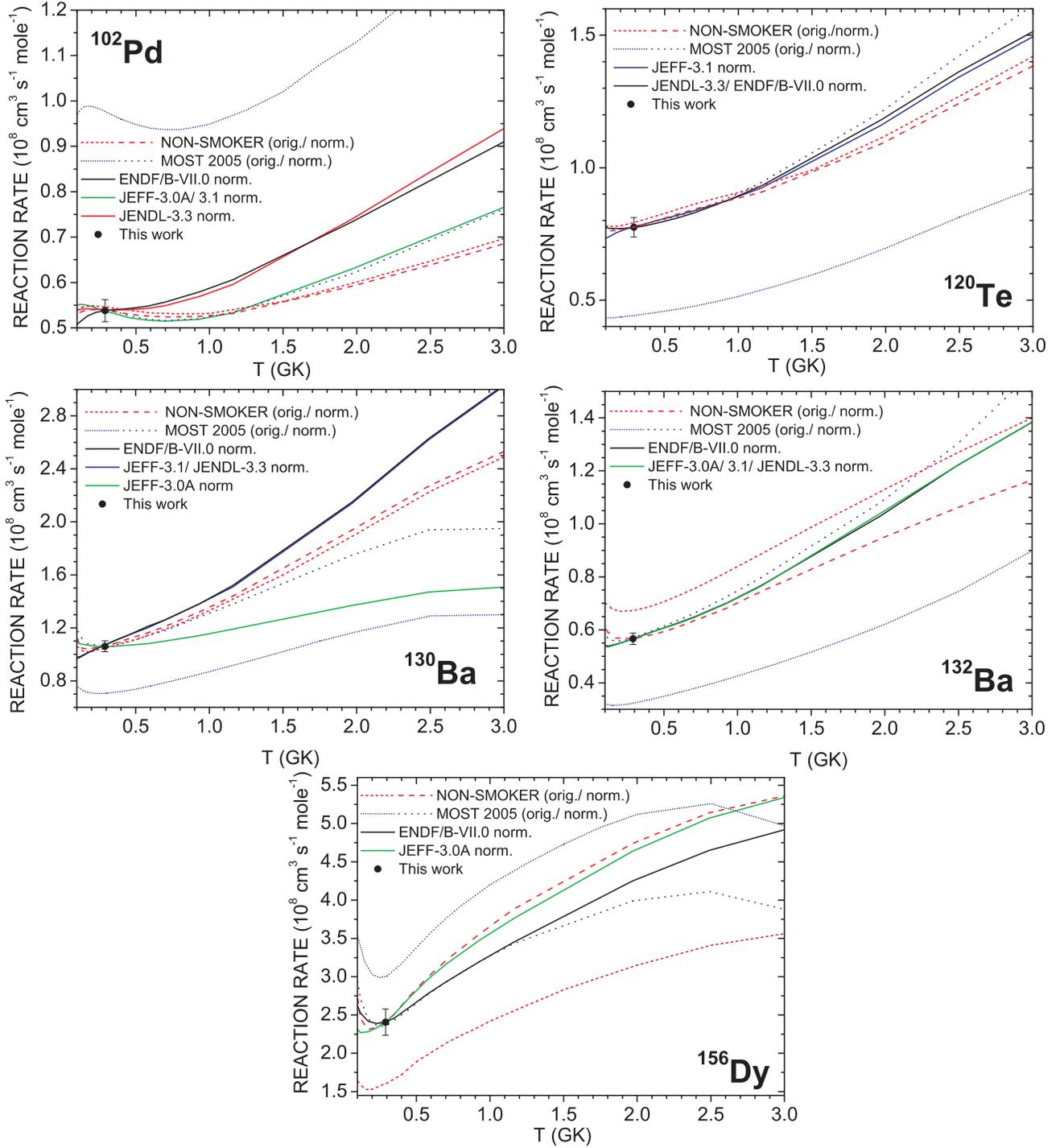}
\caption{\label{RR} (Color online) Stellar reaction rates (including f*) for temperatures between T=0.1 and 3~GK. Compared are the temperature-dependences of this work obtained with different extrapolations based on different evaluations. Also shown are the predictions of the Hauser-Feshbach models NON-SMOKER \cite{rath00,nons} and MOST 2005 \cite{most05}, each with their original values and renormalized to reproduce our data at kT=25 keV (0.29 GK). It should be noted that above 0.29 GK all values (including the ones from the evaluations) are based on energy dependences derived from Hauser-Feshbach models.}
\end{figure*}

A comparison between the normalized values from this work and the predictions of
{\sc NON-SMOKER} \cite{rath00,nons} and {\sc MOST} \cite{most05} is shown in Fig.~\ref{RR}. As can be seen and was already mentioned before, the individual evaluated data sets increasingly deviate the further the energy is from the normalization point at kT=25 keV (T=0.29 GK).

Although the {\sc MOST} results include the stellar enhancement factors, it is only a small correction at 25 keV as can be seen in Table \ref{tab:macs}. 
It has to be emphasized again that the comparison for energies $E_\mathrm{n}>25$ keV is actually a comparison with weighted Hauser-Feshbach predictions which are implicitly contained in the databases (see the discussion in Sec.\ \ref{sec:libraries}) and this way entering our extrapolation. 


Therefore, Fig.~\ref{RR} also illustrates the necessity for cross section measurements with the time-of-flight method over a wider energy range from the resonance region up to about 1 MeV, in particular for nuclei involved in the $p$-process network. Unfortunately, this will be hard to achieve, mostly because isotopically pure samples of the rare $p$-isotopes are difficult to obtain.

\section{\label{Sum}Summary and Outlook}

The (n,$\gamma$) cross sections of the $p$-isotopes $^{102}$Pd, $^{120}$Te,
$^{130,132}$Ba, and $^{156}$Dy have been measured in a quasi-stellar
neutron spectrum corresponding to a thermal energy of $kT=25$ keV by
means of the activation technique. The results for $^{102}$Pd, $^{120}$Te, 
and $^{132}$Ba represent the first experimental data, thus replacing rather 
uncertain theoretical predictions. For $^{130}$Ba and $^{156}$Dy the 
previously available experimental information could be significantly 
extended and improved. The measured cross sections were converted into 
Maxwellian averaged cross sections for a range of thermal energies
between $kT=5$ and 100 keV, and further extrapolated to the temperature
region of the $p$-process. The extrapolation still relies on theory and this
underscores the necessity for future measurements covering a wider energy range.

The present work will be complemented by a second paper on the ($n, \gamma$)
cross sections of $^{168}$Yb, $^{180}$W, $^{184}$Os, $^{190}$Pt, and $^{196}$Hg,
followed by a discussion of the astrophysical implications in a third, concluding paper.

\begin{acknowledgments}
We thank M. Brock, E. P. Knaetsch, D. Roller, and W. Seith for their help and
support during the irradiations at the Van de Graaff accelerator.
This work was supported by the Swiss National Science Foundation
Grants 2024-067428.01 and 2000-105328.
\end{acknowledgments}


\begin{thebibliography}{50}
\expandafter\ifx\csname natexlab\endcsname\relax\def\natexlab#1{#1}\fi
\expandafter\ifx\csname bibnamefont\endcsname\relax
  \def\bibnamefont#1{#1}\fi
\expandafter\ifx\csname bibfnamefont\endcsname\relax
  \def\bibfnamefont#1{#1}\fi
\expandafter\ifx\csname citenamefont\endcsname\relax
  \def\citenamefont#1{#1}\fi
\expandafter\ifx\csname url\endcsname\relax
  \def\url#1{\texttt{#1}}\fi
\expandafter\ifx\csname urlprefix\endcsname\relax\def\urlprefix{URL }\fi
\providecommand{\bibinfo}[2]{#2}
\providecommand{\eprint}[2][]{\url{#2}}

\bibitem[{\citenamefont{Burbidge et~al.}(1957)\citenamefont{Burbidge, Burbidge,
  Fowler, and Hoyle}}]{bbfh57}
\bibinfo{author}{\bibfnamefont{E.}~\bibnamefont{Burbidge}},
  \bibinfo{author}{\bibfnamefont{G.}~\bibnamefont{Burbidge}},
  \bibinfo{author}{\bibfnamefont{W.}~\bibnamefont{Fowler}}, \bibnamefont{and}
  \bibinfo{author}{\bibfnamefont{F.}~\bibnamefont{Hoyle}},
  \bibinfo{journal}{Rev. Mod. Phys.} \textbf{\bibinfo{volume}{29}},
  \bibinfo{pages}{547} (\bibinfo{year}{1957}).

\bibitem[{\citenamefont{Langanke and Wiescher}(2001)}]{lawi01}
\bibinfo{author}{\bibfnamefont{K.}~\bibnamefont{Langanke}} \bibnamefont{and}
  \bibinfo{author}{\bibfnamefont{M.}~\bibnamefont{Wiescher}},
  \bibinfo{journal}{Rep. Prog. Phys.} \textbf{\bibinfo{volume}{64}},
  \bibinfo{pages}{1657} (\bibinfo{year}{2001}).

\bibitem[{\citenamefont{Woosley and Howard}(1978)}]{woho78}
\bibinfo{author}{\bibfnamefont{S.}~\bibnamefont{Woosley}} \bibnamefont{and}
  \bibinfo{author}{\bibfnamefont{W.}~\bibnamefont{Howard}},
  \bibinfo{journal}{Astrophys. J. Suppl.} \textbf{\bibinfo{volume}{36}},
  \bibinfo{pages}{285} (\bibinfo{year}{1978}).

\bibitem[{\citenamefont{Woosley and Howard}(1990)}]{woho90}
\bibinfo{author}{\bibfnamefont{S.}~\bibnamefont{Woosley}} \bibnamefont{and}
  \bibinfo{author}{\bibfnamefont{W.}~\bibnamefont{Howard}},
  \bibinfo{journal}{Astrophys. J.} \textbf{\bibinfo{volume}{354}},
  \bibinfo{pages}{L21} (\bibinfo{year}{1990}).

\bibitem[{\citenamefont{Rayet et~al.}(1995)\citenamefont{Rayet, Arnould,
  Hashimoto, Prantzos, and Nomoto}}]{raar95}
\bibinfo{author}{\bibfnamefont{M.}~\bibnamefont{Rayet}},
  \bibinfo{author}{\bibfnamefont{M.}~\bibnamefont{Arnould}},
  \bibinfo{author}{\bibfnamefont{M.}~\bibnamefont{Hashimoto}},
  \bibinfo{author}{\bibfnamefont{N.}~\bibnamefont{Prantzos}}, \bibnamefont{and}
  \bibinfo{author}{\bibfnamefont{K.}~\bibnamefont{Nomoto}},
  \bibinfo{journal}{Astron. Astrophys.} \textbf{\bibinfo{volume}{298}},
  \bibinfo{pages}{517} (\bibinfo{year}{1995}).

\bibitem[{\citenamefont{Rauscher et~al.}(2002)\citenamefont{Rauscher, Heger,
  Hoffman, and Woosley}}]{rahe02}
\bibinfo{author}{\bibfnamefont{T.}~\bibnamefont{Rauscher}},
  \bibinfo{author}{\bibfnamefont{A.}~\bibnamefont{Heger}},
  \bibinfo{author}{\bibfnamefont{R.}~\bibnamefont{Hoffman}}, \bibnamefont{and}
  \bibinfo{author}{\bibfnamefont{S.}~\bibnamefont{Woosley}},
  \bibinfo{journal}{Astrophys. J.} \textbf{\bibinfo{volume}{576}},
  \bibinfo{pages}{323} (\bibinfo{year}{2002}).

\bibitem[{\citenamefont{Rayet et~al.}(1990)\citenamefont{Rayet, Arnould, and Prantzos}}]{ray90}
\bibinfo{author}{\bibfnamefont{M.}~\bibnamefont{Rayet}},
  \bibinfo{author}{\bibfnamefont{M.}~\bibnamefont{Arnould}}, \bibnamefont{and}
  \bibinfo{author}{\bibfnamefont{N.}~\bibnamefont{Prantzos}},
  \bibinfo{journal}{Astron. Astrophys.} \textbf{\bibinfo{volume}{227}},
  \bibinfo{pages}{271} (\bibinfo{year}{1990}).

\bibitem[{\citenamefont{Fr{\"o}hlich et~al.}(2006)\citenamefont{Fr{\"o}hlich,
  Mart\'inez-Pinedo, Liebend{\"o}rfer, Thielemann, Bravo, Hix, Langanke, and
  Zinner}}]{FML06}
\bibinfo{author}{\bibfnamefont{C.}~\bibnamefont{Fr{\"o}hlich}},
  \bibinfo{author}{\bibfnamefont{G.}~\bibnamefont{Mart\'inez-Pinedo}},
  \bibinfo{author}{\bibfnamefont{M.}~\bibnamefont{Liebend{\"o}rfer}},
  \bibinfo{author}{\bibfnamefont{F.-K.} \bibnamefont{Thielemann}},
  \bibinfo{author}{\bibfnamefont{E.}~\bibnamefont{Bravo}},
  \bibinfo{author}{\bibfnamefont{W.~R.} \bibnamefont{Hix}},
  \bibinfo{author}{\bibfnamefont{K.}~\bibnamefont{Langanke}}, \bibnamefont{and}
  \bibinfo{author}{\bibfnamefont{N.~T.} \bibnamefont{Zinner}},
  \bibinfo{journal}{Phys. Rev. Lett.} \textbf{\bibinfo{volume}{96}},
  \bibinfo{pages}{142502} (\bibinfo{year}{2006}).

\bibitem[{\citenamefont{Schatz et~al.}(1998)\citenamefont{Schatz, Aprahamian,
  G{\"o}rres, Wiescher, Rauscher, Rembges, Thielemann, Pfeiffer, M{\"o}ller,
  Herndl et~al.}}]{scha98}
\bibinfo{author}{\bibfnamefont{H.}~\bibnamefont{Schatz}},
  \bibinfo{author}{\bibfnamefont{A.}~\bibnamefont{Aprahamian}},
  \bibinfo{author}{\bibfnamefont{J.}~\bibnamefont{G{\"o}rres}},
  \bibinfo{author}{\bibfnamefont{M.}~\bibnamefont{Wiescher}},
  \bibinfo{author}{\bibfnamefont{T.}~\bibnamefont{Rauscher}},
  \bibinfo{author}{\bibfnamefont{J.}~\bibnamefont{Rembges}},
  \bibinfo{author}{\bibfnamefont{F.-K.} \bibnamefont{Thielemann}},
  \bibinfo{author}{\bibfnamefont{B.}~\bibnamefont{Pfeiffer}},
  \bibinfo{author}{\bibfnamefont{P.}~\bibnamefont{M{\"o}ller}},
  \bibinfo{author}{\bibfnamefont{H.}~\bibnamefont{Herndl}},
  \bibnamefont{et~al.}, \bibinfo{journal}{Phys. Rep.}
  \textbf{\bibinfo{volume}{294}}, \bibinfo{pages}{167} (\bibinfo{year}{1998}).

\bibitem[{\citenamefont{Schatz et~al.}(2001)\citenamefont{Schatz, Aprahamian,
  Barnard, Bildsten, Cumming, Ouellette, Rauscher, Thielemann, and
  Wiescher}}]{scha01}
\bibinfo{author}{\bibfnamefont{H.}~\bibnamefont{Schatz}},
  \bibinfo{author}{\bibfnamefont{A.}~\bibnamefont{Aprahamian}},
  \bibinfo{author}{\bibfnamefont{V.}~\bibnamefont{Barnard}},
  \bibinfo{author}{\bibfnamefont{L.}~\bibnamefont{Bildsten}},
  \bibinfo{author}{\bibfnamefont{A.}~\bibnamefont{Cumming}},
  \bibinfo{author}{\bibfnamefont{M.}~\bibnamefont{Ouellette}},
  \bibinfo{author}{\bibfnamefont{T.}~\bibnamefont{Rauscher}},
  \bibinfo{author}{\bibfnamefont{F.-K.} \bibnamefont{Thielemann}},
  \bibnamefont{and} \bibinfo{author}{\bibfnamefont{M.}~\bibnamefont{Wiescher}},
  \bibinfo{journal}{Phys. Rev. Lett.} \textbf{\bibinfo{volume}{86}},
  \bibinfo{pages}{3471} (\bibinfo{year}{2001}).

\bibitem[{\citenamefont{Woosley et~al.}(1990)\citenamefont{Woosley, Hartmann,
  Hoffman, and Haxton}}]{WHH90}
\bibinfo{author}{\bibfnamefont{S.}~\bibnamefont{Woosley}},
  \bibinfo{author}{\bibfnamefont{D.}~\bibnamefont{Hartmann}},
  \bibinfo{author}{\bibfnamefont{R.}~\bibnamefont{Hoffman}}, \bibnamefont{and}
  \bibinfo{author}{\bibfnamefont{W.}~\bibnamefont{Haxton}},
  \bibinfo{journal}{Astrophys. J.} \textbf{\bibinfo{volume}{356}},
  \bibinfo{pages}{272} (\bibinfo{year}{1990}).

\bibitem[{\citenamefont{Arnould and Goriely}(2003)}]{argo03}
\bibinfo{author}{\bibfnamefont{M.}~\bibnamefont{Arnould}} \bibnamefont{and}
  \bibinfo{author}{\bibfnamefont{S.}~\bibnamefont{Goriely}},
  \bibinfo{journal}{Phys. Rep.} \textbf{\bibinfo{volume}{384}},
  \bibinfo{pages}{1} (\bibinfo{year}{2003}).

\bibitem[{\citenamefont{Dillmann et~al.}(2006)\citenamefont{Dillmann, Heil,
  K{\"a}ppeler, Rauscher, and Thielemann}}]{DHK06}
\bibinfo{author}{\bibfnamefont{I.}~\bibnamefont{Dillmann}},
  \bibinfo{author}{\bibfnamefont{M.}~\bibnamefont{Heil}},
  \bibinfo{author}{\bibfnamefont{F.}~\bibnamefont{K{\"a}ppeler}},
  \bibinfo{author}{\bibfnamefont{T.}~\bibnamefont{Rauscher}}, \bibnamefont{and}
  \bibinfo{author}{\bibfnamefont{F.-K.} \bibnamefont{Thielemann}},
  \bibinfo{journal}{Phys. Rev. C} \textbf{\bibinfo{volume}{73}},
  \bibinfo{pages}{015803} (\bibinfo{year}{2006}).

\bibitem[{\citenamefont{Vockenhuber et~al.}(2007)\citenamefont{Vockenhuber,
  Dillmann, Heil, K{\"a}ppeler, Winckler, Kutschera, Wallner, Bichler,
  Dababneh, Bisterzo et~al.}}]{VDH07}
\bibinfo{author}{\bibfnamefont{C.}~\bibnamefont{Vockenhuber}},
  \bibinfo{author}{\bibfnamefont{I.}~\bibnamefont{Dillmann}},
  \bibinfo{author}{\bibfnamefont{M.}~\bibnamefont{Heil}},
  \bibinfo{author}{\bibfnamefont{F.}~\bibnamefont{K{\"a}ppeler}},
  \bibinfo{author}{\bibfnamefont{N.}~\bibnamefont{Winckler}},
  \bibinfo{author}{\bibfnamefont{W.}~\bibnamefont{Kutschera}},
  \bibinfo{author}{\bibfnamefont{A.}~\bibnamefont{Wallner}},
  \bibinfo{author}{\bibfnamefont{M.}~\bibnamefont{Bichler}},
  \bibinfo{author}{\bibfnamefont{S.}~\bibnamefont{Dababneh}},
  \bibinfo{author}{\bibfnamefont{S.}~\bibnamefont{Bisterzo}},
  \bibnamefont{et~al.}, \bibinfo{journal}{Phys. Rev. C}
  \textbf{\bibinfo{volume}{75}}, \bibinfo{pages}{015804}
  (\bibinfo{year}{2007}).

\bibitem[{\citenamefont{Dillmann et~al.}(2005)\citenamefont{Dillmann, Heil,
  K{\"a}ppeler, Plag, Rauscher, and Thielemann}}]{kado06}
\bibinfo{author}{\bibfnamefont{I.}~\bibnamefont{Dillmann}},
  \bibinfo{author}{\bibfnamefont{M.}~\bibnamefont{Heil}},
  \bibinfo{author}{\bibfnamefont{F.}~\bibnamefont{K{\"a}ppeler}},
  \bibinfo{author}{\bibfnamefont{R.}~\bibnamefont{Plag}},
  \bibinfo{author}{\bibfnamefont{T.}~\bibnamefont{Rauscher}}, \bibnamefont{and}
  \bibinfo{author}{\bibfnamefont{F.-K.} \bibnamefont{Thielemann}},
  \bibinfo{journal}{Proceedings of the 12th Int. Conference on Capture
  Gamma-Ray Spectroscopy and Related Topcis, Notre Dame/USA, Sept. 4-9, 2005,
  AIP Conf. Proc} \textbf{\bibinfo{volume}{819}}, \bibinfo{pages}{123}
  (\bibinfo{year}{2005}).

\bibitem[{\citenamefont{Hauser and Feshbach}(1952)}]{hafe52}
\bibinfo{author}{\bibfnamefont{W.}~\bibnamefont{Hauser}} \bibnamefont{and}
  \bibinfo{author}{\bibfnamefont{H.}~\bibnamefont{Feshbach}},
  \bibinfo{journal}{Phys. Rev.} \textbf{\bibinfo{volume}{87}},
  \bibinfo{pages}{366} (\bibinfo{year}{1952}).

\bibitem[{\citenamefont{Rauscher and Thielemann}(2000)}]{rath00}
\bibinfo{author}{\bibfnamefont{T.}~\bibnamefont{Rauscher}} \bibnamefont{and}
  \bibinfo{author}{\bibfnamefont{F.-K.} \bibnamefont{Thielemann}},
  \bibinfo{journal}{At. Data Nucl. Data Tables} \textbf{\bibinfo{volume}{75}},
  \bibinfo{pages}{1} (\bibinfo{year}{2000}).

\bibitem[{\citenamefont{Rauscher and Thielemann}(2001)}]{rath01}
\bibinfo{author}{\bibfnamefont{T.}~\bibnamefont{Rauscher}} \bibnamefont{and}
  \bibinfo{author}{\bibfnamefont{F.-K.} \bibnamefont{Thielemann}},
  \bibinfo{journal}{At. Data Nucl. Data Tables} \textbf{\bibinfo{volume}{79}},
  \bibinfo{pages}{47} (\bibinfo{year}{2001}).

\bibitem[{\citenamefont{Goriely}(2005)}]{most05}
\bibinfo{author}{\bibfnamefont{S.}~\bibnamefont{Goriely}},
  \bibinfo{journal}{Hauser-Feshbach rates for neutron capture reactions
  (version 08/26/05), http://www-astro.ulb.ac.be/Html/hfr.html}
  (\bibinfo{year}{2005}).

\bibitem[{\citenamefont{Beer and K{\"a}ppeler}(1980)}]{beer80}
\bibinfo{author}{\bibfnamefont{H.}~\bibnamefont{Beer}} \bibnamefont{and}
  \bibinfo{author}{\bibfnamefont{F.}~\bibnamefont{K{\"a}ppeler}},
  \bibinfo{journal}{Phys. Rev. C} \textbf{\bibinfo{volume}{21}},
  \bibinfo{pages}{534} (\bibinfo{year}{1980}).

\bibitem[{\citenamefont{Ratynski and K{\"a}ppeler}(1988)}]{raty88}
\bibinfo{author}{\bibfnamefont{W.}~\bibnamefont{Ratynski}} \bibnamefont{and}
  \bibinfo{author}{\bibfnamefont{F.}~\bibnamefont{K{\"a}ppeler}},
  \bibinfo{journal}{Phys. Rev. C} \textbf{\bibinfo{volume}{37}},
  \bibinfo{pages}{595} (\bibinfo{year}{1988}).

\bibitem[{\citenamefont{De~Laeter et~al.}(2003)\citenamefont{De~Laeter,
  B{\"o}hlke, de~Bievre, Hidaka, Peiser, Rosman, and Taylor}}]{iupac03}
\bibinfo{author}{\bibfnamefont{J.}~\bibnamefont{De~Laeter}},
  \bibinfo{author}{\bibfnamefont{J.}~\bibnamefont{B{\"o}hlke}},
  \bibinfo{author}{\bibfnamefont{P.}~\bibnamefont{de~Bievre}},
  \bibinfo{author}{\bibfnamefont{H.}~\bibnamefont{Hidaka}},
  \bibinfo{author}{\bibfnamefont{H.}~\bibnamefont{Peiser}},
  \bibinfo{author}{\bibfnamefont{K.}~\bibnamefont{Rosman}}, \bibnamefont{and}
  \bibinfo{author}{\bibfnamefont{P.}~\bibnamefont{Taylor}},
  \bibinfo{journal}{Pure and Appl. Chem.} \textbf{\bibinfo{volume}{75}},
  \bibinfo{pages}{683} (\bibinfo{year}{2003}).

\bibitem[{\citenamefont{Smith et~al.}(1978)\citenamefont{Smith, Rosman, and
  De~Laeter}}]{SRL78}
\bibinfo{author}{\bibfnamefont{C.}~\bibnamefont{Smith}},
  \bibinfo{author}{\bibfnamefont{K.}~\bibnamefont{Rosman}}, \bibnamefont{and}
  \bibinfo{author}{\bibfnamefont{J.}~\bibnamefont{De~Laeter}},
  \bibinfo{journal}{Int. J. Mass Spectrom. Ion Phys.}
  \textbf{\bibinfo{volume}{28}}, \bibinfo{pages}{7} (\bibinfo{year}{1978}).

\bibitem[{\citenamefont{De~Laeter}(1994)}]{deLae94}
\bibinfo{author}{\bibfnamefont{J.}~\bibnamefont{De~Laeter}},
  \bibinfo{journal}{Astrophys. J.} \textbf{\bibinfo{volume}{434}},
  \bibinfo{pages}{695} (\bibinfo{year}{1994}).

\bibitem[{\citenamefont{Lee and Halliday}(1995)}]{Lee95}
\bibinfo{author}{\bibfnamefont{D.-C.} \bibnamefont{Lee}} \bibnamefont{and}
  \bibinfo{author}{\bibfnamefont{A.}~\bibnamefont{Halliday}},
  \bibinfo{journal}{Int. J. Mass Spectrom. Ion Proc.}
  \textbf{\bibinfo{volume}{146/147}}, \bibinfo{pages}{35}
  (\bibinfo{year}{1995}).

\bibitem[{\citenamefont{Dababneh et~al.}(2004)\citenamefont{Dababneh, Patronis,
  Assimakopoulos, G{\"o}rres, Heil, K{\"a}ppeler, Karamanis, O'Brien, and
  Reifarth}}]{clover}
\bibinfo{author}{\bibfnamefont{S.}~\bibnamefont{Dababneh}},
  \bibinfo{author}{\bibfnamefont{N.}~\bibnamefont{Patronis}},
  \bibinfo{author}{\bibfnamefont{P.}~\bibnamefont{Assimakopoulos}},
  \bibinfo{author}{\bibfnamefont{J.}~\bibnamefont{G{\"o}rres}},
  \bibinfo{author}{\bibfnamefont{M.}~\bibnamefont{Heil}},
  \bibinfo{author}{\bibfnamefont{F.}~\bibnamefont{K{\"a}ppeler}},
  \bibinfo{author}{\bibfnamefont{D.}~\bibnamefont{Karamanis}},
  \bibinfo{author}{\bibfnamefont{S.}~\bibnamefont{O'Brien}}, \bibnamefont{and}
  \bibinfo{author}{\bibfnamefont{R.}~\bibnamefont{Reifarth}},
  \bibinfo{journal}{Nucl. Instr. and Meth. A} \textbf{\bibinfo{volume}{517}},
  \bibinfo{pages}{230} (\bibinfo{year}{2004}).

\bibitem[{\citenamefont{de~Frenne and Jacobs}(2001)}]{nds103}
\bibinfo{author}{\bibfnamefont{D.}~\bibnamefont{de~Frenne}} \bibnamefont{and}
  \bibinfo{author}{\bibfnamefont{E.}~\bibnamefont{Jacobs}},
  \bibinfo{journal}{Nucl. Data Sheets} \textbf{\bibinfo{volume}{93}},
  \bibinfo{pages}{447} (\bibinfo{year}{2001}).

\bibitem[{\citenamefont{Tamura}(2000)}]{nds121}
\bibinfo{author}{\bibfnamefont{T.}~\bibnamefont{Tamura}},
  \bibinfo{journal}{Nucl. Data Sheets} \textbf{\bibinfo{volume}{90}},
  \bibinfo{pages}{107} (\bibinfo{year}{2000}).

\bibitem[{\citenamefont{Khazov et~al.}(2006)\citenamefont{Khazov, Mitropolsky,
  and Rodionov}}]{nds131new}
\bibinfo{author}{\bibfnamefont{Y.}~\bibnamefont{Khazov}},
  \bibinfo{author}{\bibfnamefont{I.}~\bibnamefont{Mitropolsky}},
  \bibnamefont{and} \bibinfo{author}{\bibfnamefont{A.}~\bibnamefont{Rodionov}},
  \bibinfo{journal}{Nucl. Data Sheets} \textbf{\bibinfo{volume}{1072}},
  \bibinfo{pages}{2715} (\bibinfo{year}{2006}).

\bibitem[{\citenamefont{Shaheen}(1995)}]{nds133a}
\bibinfo{author}{\bibfnamefont{R.}~\bibnamefont{Shaheen}},
  \bibinfo{journal}{Nucl. Data Sheets} \textbf{\bibinfo{volume}{75}},
  \bibinfo{pages}{491} (\bibinfo{year}{1995}).

\bibitem[{\citenamefont{Helmer}(2004)}]{nds157}
\bibinfo{author}{\bibfnamefont{R.}~\bibnamefont{Helmer}},
  \bibinfo{journal}{Nucl. Data Sheets} \textbf{\bibinfo{volume}{103}},
  \bibinfo{pages}{565} (\bibinfo{year}{2004}).

\bibitem[{\citenamefont{Chunmei}(2002)}]{nds198}
\bibinfo{author}{\bibfnamefont{Z.}~\bibnamefont{Chunmei}},
  \bibinfo{journal}{Nucl. Data Sheets} \textbf{\bibinfo{volume}{95}},
  \bibinfo{pages}{59} (\bibinfo{year}{2002}).

\bibitem[{\citenamefont{Hubbell and Seltzer}(2004)}]{nist}
\bibinfo{author}{\bibfnamefont{J.}~\bibnamefont{Hubbell}} \bibnamefont{and}
  \bibinfo{author}{\bibfnamefont{S.}~\bibnamefont{Seltzer}},
  \bibinfo{journal}{Table of X-Ray Mass Attenuation Coefficients and Mass
  Energy-Absorption Coefficients (v. 1.4), National Institute of Standards and
  Technology, Gaithersburg, MD ;
  http://physics.nist.gov/PhysRefData/XrayMassCoef/}  (\bibinfo{year}{2004}).

\bibitem[{\citenamefont{Agostinelli et~al.}(2003)\citenamefont{Agostinelli,
  Allison, Amako, Apostolakis, Araujo, Arce, Asai, Axen, Banerjee, Barrand
  et~al.}}]{Ago03}
\bibinfo{author}{\bibfnamefont{S.}~\bibnamefont{Agostinelli}},
  \bibinfo{author}{\bibfnamefont{J.}~\bibnamefont{Allison}},
  \bibinfo{author}{\bibfnamefont{K.}~\bibnamefont{Amako}},
  \bibinfo{author}{\bibfnamefont{J.}~\bibnamefont{Apostolakis}},
  \bibinfo{author}{\bibfnamefont{H.}~\bibnamefont{Araujo}},
  \bibinfo{author}{\bibfnamefont{P.}~\bibnamefont{Arce}},
  \bibinfo{author}{\bibfnamefont{M.}~\bibnamefont{Asai}},
  \bibinfo{author}{\bibfnamefont{D.}~\bibnamefont{Axen}},
  \bibinfo{author}{\bibfnamefont{S.}~\bibnamefont{Banerjee}},
  \bibinfo{author}{\bibfnamefont{G.}~\bibnamefont{Barrand}},
  \bibnamefont{et~al.}, \bibinfo{journal}{Nucl. Instr. Meth. A}
  \textbf{\bibinfo{volume}{506}}, \bibinfo{pages}{250} (\bibinfo{year}{2003}).

\bibitem[{\citenamefont{Mughabghab et~al.}(1981)\citenamefont{Mughabghab,
  Divadeenam, and Holden}}]{mugh81}
\bibinfo{author}{\bibfnamefont{S.}~\bibnamefont{Mughabghab}},
  \bibinfo{author}{\bibfnamefont{M.}~\bibnamefont{Divadeenam}},
  \bibnamefont{and} \bibinfo{author}{\bibfnamefont{N.}~\bibnamefont{Holden}},
  \bibinfo{journal}{Neutron Cross Sections, BNL-325, 1st ed.,}
  \textbf{\bibinfo{volume}{1}} (\bibinfo{year}{1981}).

\bibitem[{\citenamefont{Mughabghab}(2006)}]{mugh06}
\bibinfo{author}{\bibfnamefont{S.}~\bibnamefont{Mughabghab}},
  \bibinfo{journal}{Atlas of Neutron Resonances (5th ed.), Elsevier, ISBN
  0-444-52035-X}  (\bibinfo{year}{2006}).

\bibitem[{\citenamefont{Soppera et~al.}(2008)\citenamefont{Soppera, Bossant,
  Henriksson, Nagel, and Rugama}}]{janis30}
\bibinfo{author}{\bibfnamefont{N.}~\bibnamefont{Soppera}},
  \bibinfo{author}{\bibfnamefont{M.}~\bibnamefont{Bossant}},
  \bibinfo{author}{\bibfnamefont{H.}~\bibnamefont{Henriksson}},
  \bibinfo{author}{\bibfnamefont{P.}~\bibnamefont{Nagel}}, \bibnamefont{and}
  \bibinfo{author}{\bibfnamefont{Y.}~\bibnamefont{Rugama}},
  \bibinfo{journal}{International Conference on Nuclear Data for Science and
  Technology, Nice/ France 2007, edts. O. Bersillon, F. Gunsing, E. Bauge, R.
  Jacqmin, S. Leray, EDP Sciences, ISBN: 978-2-7598-0091-9} p.
  \bibinfo{pages}{773} (\bibinfo{year}{2008}).

\bibitem[{\citenamefont{Shibata et~al.}(2002)\citenamefont{Shibata, Kawano, and
  Nakagawa}}]{jendl33}
\bibinfo{author}{\bibfnamefont{K.}~\bibnamefont{Shibata}},
  \bibinfo{author}{\bibfnamefont{T.}~\bibnamefont{Kawano}}, \bibnamefont{and}
  \bibinfo{author}{\bibfnamefont{T.}~\bibnamefont{Nakagawa}},
  \bibinfo{journal}{Japanese Evaluated Nuclear Data Library Version 3 Revision
  3: JENDL 3.3, J. Nucl. Sci. Technol.} \textbf{\bibinfo{volume}{39}},
  \bibinfo{pages}{1125} (\bibinfo{year}{2002}).

\bibitem[{\citenamefont{Chadwick et~al.}(2006)\citenamefont{Chadwick,
  Oblozinsk�, Herman, Greene, McKnight, Smith, Young, MacFarlane, Hale,
  Frankle et~al.}}]{endfb7}
\bibinfo{author}{\bibfnamefont{M.}~\bibnamefont{Chadwick}},
  \bibinfo{author}{\bibfnamefont{P.}~\bibnamefont{Oblozinsk�}},
  \bibinfo{author}{\bibfnamefont{M.}~\bibnamefont{Herman}},
  \bibinfo{author}{\bibfnamefont{N.}~\bibnamefont{Greene}},
  \bibinfo{author}{\bibfnamefont{R.}~\bibnamefont{McKnight}},
  \bibinfo{author}{\bibfnamefont{D.}~\bibnamefont{Smith}},
  \bibinfo{author}{\bibfnamefont{P.}~\bibnamefont{Young}},
  \bibinfo{author}{\bibfnamefont{R.}~\bibnamefont{MacFarlane}},
  \bibinfo{author}{\bibfnamefont{G.}~\bibnamefont{Hale}},
  \bibinfo{author}{\bibfnamefont{S.}~\bibnamefont{Frankle}},
  \bibnamefont{et~al.}, \bibinfo{journal}{Nuclear Data Sheets}
  \textbf{\bibinfo{volume}{107}}, \bibinfo{pages}{2931} (\bibinfo{year}{2006}).

\bibitem[{\citenamefont{Igarasi and Fukahori}(1991)}]{casthy}
\bibinfo{author}{\bibfnamefont{S.}~\bibnamefont{Igarasi}} \bibnamefont{and}
  \bibinfo{author}{\bibfnamefont{T.}~\bibnamefont{Fukahori}},
  \bibinfo{journal}{JAERI 1321}  (\bibinfo{year}{1991}).

\bibitem[{\citenamefont{Bradley et~al.}(1979)\citenamefont{Bradley, Parsa,
  Stelts, and Chrien}}]{BPS79}
\bibinfo{author}{\bibfnamefont{T.}~\bibnamefont{Bradley}},
  \bibinfo{author}{\bibfnamefont{Z.}~\bibnamefont{Parsa}},
  \bibinfo{author}{\bibfnamefont{M.}~\bibnamefont{Stelts}}, \bibnamefont{and}
  \bibinfo{author}{\bibfnamefont{R.}~\bibnamefont{Chrien}},
  \bibinfo{journal}{\emph{Nuclear Cross Sections for Technology}, edt. by J.L.
  Fowler, C.H. Johnson, and C.D. Bowman (National Bureau of Standards,
  Washington D.C.)} p. \bibinfo{pages}{344} (\bibinfo{year}{1979}).

\bibitem[{\citenamefont{Young et~al.}(1998)\citenamefont{Young, Arthur, and
  Chadwick}}]{gnash}
\bibinfo{author}{\bibfnamefont{P.}~\bibnamefont{Young}},
  \bibinfo{author}{\bibfnamefont{V.}~\bibnamefont{Arthur}}, \bibnamefont{and}
  \bibinfo{author}{\bibfnamefont{M.}~\bibnamefont{Chadwick}},
  \bibinfo{journal}{\textit{Proceedings of the IAEA Workshop on Nuclear
  Reaction Data and Nuclear Reactors: Physics, Design and Safety}, edt. by A.
  Gandini and G. Reffo, Trieste/Italy, 1996, World Scientific Publishing, Ltd.,
  Singapore} p. \bibinfo{pages}{227} (\bibinfo{year}{1998}).

\bibitem[{\citenamefont{{Herman} et~al.}(2007)\citenamefont{{Herman}, {Capote},
  {Carlson}, {Oblo\v{z}insk\'{y}}, {Sin}, {Trkov}, {Wienke}, and
  {Zerkin}}}]{empire}
\bibinfo{author}{\bibfnamefont{M.}~\bibnamefont{{Herman}}},
  \bibinfo{author}{\bibfnamefont{R.}~\bibnamefont{{Capote}}},
  \bibinfo{author}{\bibfnamefont{B.}~\bibnamefont{{Carlson}}},
  \bibinfo{author}{\bibfnamefont{P.}~\bibnamefont{{Oblo\v{z}insk\'{y}}}},
  \bibinfo{author}{\bibfnamefont{M.}~\bibnamefont{{Sin}}},
  \bibinfo{author}{\bibfnamefont{A.}~\bibnamefont{{Trkov}}},
  \bibinfo{author}{\bibfnamefont{H.}~\bibnamefont{{Wienke}}}, \bibnamefont{and}
  \bibinfo{author}{\bibfnamefont{V.}~\bibnamefont{{Zerkin}}},
  \bibinfo{journal}{Nucl. Data Sheets} \textbf{\bibinfo{volume}{108}},
  \bibinfo{pages}{2655} (\bibinfo{year}{2007}).

\bibitem[{\citenamefont{Rauscher}(2009)}]{nons}
\bibinfo{author}{\bibfnamefont{T.}~\bibnamefont{Rauscher}},
  \bibinfo{journal}{HTML Interface to NON-SMOKER; Online:
  http://nucastro.org/nonsmoker}  (\bibinfo{year}{2009}).

\bibitem[{\citenamefont{Bao et~al.}(2000)\citenamefont{Bao, Beer, K{\"a}ppeler,
  Voss, Wisshak, and Rauscher}}]{bao00}
\bibinfo{author}{\bibfnamefont{Z.}~\bibnamefont{Bao}},
  \bibinfo{author}{\bibfnamefont{H.}~\bibnamefont{Beer}},
  \bibinfo{author}{\bibfnamefont{F.}~\bibnamefont{K{\"a}ppeler}},
  \bibinfo{author}{\bibfnamefont{F.}~\bibnamefont{Voss}},
  \bibinfo{author}{\bibfnamefont{K.}~\bibnamefont{Wisshak}}, \bibnamefont{and}
  \bibinfo{author}{\bibfnamefont{T.}~\bibnamefont{Rauscher}},
  \bibinfo{journal}{At. Data Nucl. Data Tables} \textbf{\bibinfo{volume}{76}},
  \bibinfo{pages}{70} (\bibinfo{year}{2000}).

\bibitem[{\citenamefont{Allen et~al.}(1971)\citenamefont{Allen, Gibbons, and
  Macklin}}]{alle71}
\bibinfo{author}{\bibfnamefont{B.}~\bibnamefont{Allen}},
  \bibinfo{author}{\bibfnamefont{J.}~\bibnamefont{Gibbons}}, \bibnamefont{and}
  \bibinfo{author}{\bibfnamefont{R.}~\bibnamefont{Macklin}},
  \bibinfo{journal}{Adv. Nucl. Phys.} \textbf{\bibinfo{volume}{4}},
  \bibinfo{pages}{205} (\bibinfo{year}{1971}).

\bibitem[{\citenamefont{Beer}(1985)}]{Bee85}
\bibinfo{author}{\bibfnamefont{H.}~\bibnamefont{Beer}},
  \bibinfo{journal}{Technical report KfK-3969, Kernforschungszentrum Karlsruhe}
  p.~\bibinfo{pages}{14} (\bibinfo{year}{1985}).

\bibitem[{\citenamefont{Holmes et~al.}(1976)\citenamefont{Holmes, Woosley,
  Fowler, and Zimmerman}}]{holm76}
\bibinfo{author}{\bibfnamefont{J.}~\bibnamefont{Holmes}},
  \bibinfo{author}{\bibfnamefont{S.}~\bibnamefont{Woosley}},
  \bibinfo{author}{\bibfnamefont{W.}~\bibnamefont{Fowler}}, \bibnamefont{and}
  \bibinfo{author}{\bibfnamefont{B.}~\bibnamefont{Zimmerman}},
  \bibinfo{journal}{At. Data Nucl. Data Tables} \textbf{\bibinfo{volume}{18}},
  \bibinfo{pages}{305} (\bibinfo{year}{1976}).

\bibitem[{\citenamefont{Harris}(1981)}]{harr81}
\bibinfo{author}{\bibfnamefont{M.}~\bibnamefont{Harris}},
  \bibinfo{journal}{Astrophys. Space Sci.} \textbf{\bibinfo{volume}{77}},
  \bibinfo{pages}{357} (\bibinfo{year}{1981}).

\bibitem[{\citenamefont{Zhao et~al.}(1988)\citenamefont{Zhao, Zhou, and
  Cai}}]{zhao88}
\bibinfo{author}{\bibfnamefont{Z.}~\bibnamefont{Zhao}},
  \bibinfo{author}{\bibfnamefont{D.}~\bibnamefont{Zhou}}, \bibnamefont{and}
  \bibinfo{author}{\bibfnamefont{D.}~\bibnamefont{Cai}},
  \bibinfo{journal}{Nuclear Data for Science and Technology, edts. S. Igarasi
  (Saikon, Tokyo)} p. \bibinfo{pages}{513} (\bibinfo{year}{1988}).

\end{thebibliography}

\end{document}